\documentclass[fleqn,usenatbib]{mnras}

\usepackage{newtxtext,newtxmath}

\usepackage[T1]{fontenc}
\usepackage{ae,aecompl}

\usepackage{graphicx}	
\usepackage{amsmath}	
\usepackage{amssymb}	
\usepackage{gensymb}
\usepackage{natbib}
\usepackage{float}
\usepackage{subfig}
\usepackage[section]{placeins}
\usepackage{amsmath}
\newcommand{\arcdeg}{\ensuremath{^{\circ}}}%

\title[The complex jet- and bar-perturbed kinematics in NGC 3393]{The complex jet- and bar-perturbed kinematics in NGC 3393 as revealed with ALMA and GEMINI-GMOS/IFU}

\author[Finlez, Carolina et al.]{
Finlez, Carolina,$^{1}$\thanks{E-mail: cfinlez@udec.cl (CF)}
Nagar, Neil M.$^{1}$
Storchi-Bergmann, Thaisa$^{2}$ 
Schnorr-M\"{u}ller, Allan$^{2}$
\newauthor
Riffel, Rogemar A.$^{3}$
Lena, Davide$^{4,5}$
Mundell, C. G. $^{6}$
Elvis, Martin S. $^{7}$
\\
$^{1}$Departamento de Astronom\'ia, Universidad de Concepci\'on, Casilla 160-C, Concepci\'on, Chile \\
$^{2}$Instituto de F\'isica, Universidade Federal do Rio Grande do Sul, 91501-970 Porto Alegre, RS, Brazil \\
$^{3}$Universidade Federal de Santa Maria, Departamento de F\'isica, Centro de Ciencias Naturales e Exatas, 97105-900, Santa Maria, RS, Brazil\\
$^{4}$SRON Netherlands Institute for Space Research, Sorbonnelaan 2, NL-3584 CA Utretch, the Netherlands\\
$^{5}$Department of Astrophysics/IMAPP, Radboud University, Nijmegen, PO Box 9010, NL-6500 GL Nijmegen, the Netherlands\\
$^{6}$Department of Physics, University of Bath, Claverton Down, Bath, BA2 7AY, UK\\
$^{7}$Harvard-Smithsonian Center for Astrophysics, 60 Garden St., Cambridge, MA 02138, USA
}

\date{Accepted XXX. Received YYY; in original form ZZZ}

\pubyear{2'18}

\begin{document}
\label{firstpage}
\pagerange{\pageref{firstpage}--\pageref{lastpage}}
\maketitle

\begin{abstract}
NGC 3393, a nearby Seyfert 2 galaxy with nuclear radio jets, large-scale and nuclear bars, and a posited secondary super massive black hole, provides an interesting laboratory to test the physics of inflows and outflows. Here we present and analyse the molecular gas (ALMA observations of CO J:2-1 emission over a field of view (FOV) of 45\arcsec $\times$ 45\arcsec, at 0\farcs56 (143 pc) spatial and 5 km/s spectral resolution), ionised gas and stars (GEMINI-GMOS/IFU; over a FOV of 4\arcsec $\times$ 5\arcsec, at 0\farcs62 (159 pc) spatial and 23 km/s spectral resolution) in NGC 3393.
The ionised gas emission, detected over the complete GEMINI-GMOS FOV, has three identifiable kinematic components. A narrow ($\sigma <$ 115 km/s) component present in the complete FOV, which is consistent with rotation in the galaxy disk. A broad ($\sigma >$ 115 km/s) redshifted component, detected near the NE and SW radio lobes; which we interpret as a radio jet driven outflow. And a broad ($\sigma >$ 115 km/s) blueshifted component that shows high velocities in a region perpendicular to the radio jet axis; we interpret this as an equatorial outflow. 
The CO J:2-1 emission is detected in spiral arms on 5\arcsec $-$ 20\arcsec scales, and in two disturbed circumnuclear regions. The molecular kinematics in the spiral arms can be explained by rotation. The highly disturbed kinematics of the inner region can be explained by perturbations induced by the nuclear bar and interactions with the large scale bar.
 We find no evidence for, but cannot strongly rule out, the presence of the posited secondary black hole.
\end{abstract}

\begin{keywords}
black hole physics - galaxies: active - galaxies: individual (NGC 3393) - galaxies: kinematics and dynamics - galaxies: nuclei - galaxies: Seyfert
\end{keywords}



\section{Introduction}
\label{sect:introduction}

It is widely accepted that most -- if not all -- nearby galaxies with a bulge component host a super massive black hole (SMBH) in their centre \citep[e.g.][]{ferrarese+merrit2000,kormendy+richstone1995,ferrarese+ford2005}. A considerable amount of observational evidence supports a connection between SMBH and the host galaxy growth, the main one being a strong correlation between the mass of the SMBH and the properties of the bulge of the host galaxy \citep{gebhardt2000}; nevertheless, the details of coevolution of the SMBH and the host galaxy remain the subject of an ongoing debate.

Despite the ubiquity of SMBHs at the centre of galaxies, only a small fraction of these are active galactic nuclei (AGN) in the local universe  \citep[e.g.][]{kewley2006,ho+filippenko1997}. The questions of what triggers activity in a galactic centre and if this ignition mechanism is related to the host galaxy properties, arise naturally. The lack of activity in the nucleus can be related to a lack of accretion material or the absence of a fuelling mechanism. The amount of gas needed to fuel an AGN over a normal duty cycle is a large fraction of the total gas contained in the galaxy \citep{combes2001}, thus a fuelling mechanism must be able to remove most of the angular momentum of a large amount of gas so this can be transferred from the kpc-scale into the sub-pc central region to feed the AGN.

Gravitational mechanisms such as galaxy interactions can drive gas inwards. Major mergers are thought to be responsible for the high accretion rate observed in luminous quasars, while minor mergers can produce Seyfert-level luminosities. Alternatively, secular mechanisms can also remove angular momentum and drive gas towards the nucleus. Observations point to secular processes being the most common triggering mechanisms for medium to low luminosity AGNs. \citep[e.g.][]{hopkins2014,treister2012,fan2016,goulding2017b}

Non-axisymmetric potentials, such as spiral structures and bars can produce radial inflows to the central region, which can be observed as line-of-sight velocity distortions \citep{lin+shu1964,lindblad1964}. Gas transport via bars is efficient \citep[e.g.][]{mundel+shone1999} from large scale down to the inner kpc, where gas can get stalled in rings at the inner Lindblad resonance region \citep{combes+gerin1985}. From this scale to the centre other mechanisms can be invoked to transport gas to the AGN, such as inner spiral structures and bars within bars \citep{shlosman1989}. \\

The incidence of bars is similar in both active and non-active galaxies and thus a strong correlation between bar presence and activity is yet to be found \citep[e.g.][]{knapen2000,cisternas2013,galloway2015,cheung2015,goulding2017a} .
However a difference between active and inactive galaxies has been observed by \citet{simoeslopes2007} and \citet{martini+pogge1999}. Using a sample of active and control galaxy pairs, they observed nuclear dust spirals and structures in 100\% of the active early-type galaxies, but in only 25\% of the control sample. This dust excess has been confirmed by \citet{martini+dicken2013}, and is thought to trace the feeding channels to the AGN \citep{ferrarese+ford2005,kormendy+ho2013,storchibergmann2007}. 
Similarly, in a study of ionised gas dynamics in a matched sample of active and inactive galaxies, \citet{dumas2007}, identified increased kinematics disturbance as a function of accretion rate in the inner 1 kpc of AGN, where activity and dynamical timescales become comparable. \\

The non-axisymmetric gravitational potential created by a bar can produce important kinematic effects on the gas which have been studied by hydrodynamical \citep[e.g][]{lindblad+lindblad1996, kim2012, athanassoula1992} and/or N-body simulations \citep[e.g.][]{sellwood1981,emsellem2001a}. 
A different method to gain insight on these kinematic effects is to quantify the line-of-sight deviations from pure rotation, which can be achieved from linear perturbation theory \citep{lin+shu1964}. The non-axisymmetric distortions to the planar flow can be decomposed in their harmonic components. \citet{franx1994} and \citet{schoenmakers1997} pioneered an approach to interpret these harmonic coefficients based on epicycle theory, and thus only valid for small departures from circular orbit speed. Since then, this harmonic decomposition analysis has been carried by several authors, e.g., \citet{wong2004,emsellem2001b}. \\

One aspect of the SMBH-host galaxy connection is the interaction between the gas of the host galaxy and the energy generated by the AGN, which produces a feedback process that has been theorised as an important component in galaxy evolution as it can help to regulate the growth of the galaxy, preventing it from becoming too massive \citep{dimatteo2005, wagner+bicknell2011,fabian2012}. \\
This interaction can occur, broadly, in two main modes: radiative or quasar mode and kinetic or radio mode \citep{croton2006,fabian2012}. The former is driven by a wind caused by the accretion of material into the SMBH, producing wide-angle sub-relativistic outflows. The latter is driven by relativistic radio jets.
Both these winds and jets can have important consequences in the galaxy evolution as they can heat and ionise cold gas when colliding with it, preventing the gas from collapsing under self-gravity, thus halting the accretion onto the SMBH and quenching star formation \citep[e.g.][]{hardcastle2013,best2005}. The jets can also directly expel gas from the galaxy removing the components for further star formation \citep{nesvadba2006}. However, some simulations \citep[e.g.][]{antonuccio-delogu2010,silk+nusser2010} reveal that jet activity is able to trigger star formation by producing high density cavities with low temperature, which are embedded in a cocoon around the jet \citep[e.g.][]{best1998,jarvis2001}, a scenario known as positive radio mode feedback. 
The alignment of the outflowing gas with the jet suggests that the outflows are driven by the transfer of energy and momentum from the radio jet to the ISM, as shown by hydrodynamical simulations \citep{wagner2012}. Jet-driven outflows have been observed in neutral and molecular gas \citep[e.g.][]{morganti2005,dasyra+combes2012} and in the ionised gas \citep[e.g.][]{holt2011,riffel2006,couto2017}. Kinematic features consistent with gas inflows and outflows have been found in the ionised and molecular gas of the central region of nearby galaxies \citep[e.g.][and references therein]{lena2016,schnorrmuller2016,schnorrmuller2017,garciaburillo2005}. \\

In this work we analyse the bar- and jet- induced perturbations on the molecular and ionised gas in NGC 3393 using new data from ALMA (CO J:2-1) and GEMINI-GMOS/IFU (optical spectra with information from stars and gas). 
NGC 3393 is a nearby, bright \citep[$m_{b}=13.1$ according to][]{devaucouleurs1991}, spiral (Sa) galaxy, at an estimated  redshift of 0.012442 (optical), or 0.012509 (from the 21 cm line) which corresponds to a luminosity distance of $52$ Mpc and a scale of $0.25$ kpc/arcsec, assuming  $H_{o} = 73$ km s$^{-1}$ Mpc$^{-1}$. The galaxy covers over one arcmin on the sky, it is observed nearly face-on, and it has been classified optically as a Seyfert 2 \citep{veroncetty2003}. It is interacting weakly with a nearby companion that is 60 kpc away \citep{schmitt2001}.

 From HI single dish observations the maximum rotation velocity corrected by inclination is $158 \pm 7$ km/s, and the central velocity dispersion is $197 \pm 28 $km/s  \citep{leda-paturel2003}.  
NIR images show a stellar bar in PA $\sim 159 \degree - 165 \degree$, with maximum ellipticity $e_{max} = 0.2$, and semi major axis (SMA) $13 \arcsec$. A faint nuclear bar has also been posited in position angle (PA) $\sim 145\degree - 150\degree$, $e_{max}=0.46$, and SMA  $2 \arcsec$ \citep{alonso-herrero1998,jungwiert1997}.
\citet{lasker2016} modelled the light distribution of the galaxy using HST imaging and found two prominent rings: the first is actually a partial ring formed by two asymmetric tightly-wound spiral arms in an outer disk of radius $40 \arcsec$; the second is an inner ring that appears elongated, with a radius of $13\arcsec$, i.e. coincident with the outer bar identified by \citet{alonso-herrero1998}. These authors derive a PA of $140\degree$ for the $2\arcsec$ SMA inner bar.

A high resolution HST [OIII] emission line image of the NLR \citep{Schmitt2003} shows an S-shaped morphology with arms that show an opening angle of $90 \degree $, and an extension of $5.6\arcsec$ pc (1410 pc along the ionisation axis PA) $\times\ 3\arcsec$ (740 pc), with the ionisation axis oriented in PA $65\degree$. Their derived PA is only slightly different from the value of $55\degree$ quoted by \citet{schmitt1996} and \citet{cooke2000}. 
The sense of curvature of this S-shape is the same as the large-scale spiral arms. The [OIII] emission extends up to $r \sim 15\arcsec$ (3750 pc) along PA $44 \degree$. 
This S-shaped structure of high-excitation gas surrounds a three-component radio structure, as observed with the VLA at 1.5, 4 and 8.4 GHz \citep{koss2015,cooke2000}. The central radio source is unresolved and has a flatter spectrum than the lobes. Chandra data was also obtained by \citet{bianchi2006} and \citet{levenson2006}, who found soft X-ray emission that has strong morphological correlations with the extended [OIII] emission.\\
The kinematics of the NLR has been studied by \citet{cooke2000} using Fabry-P\'erot [NII] data. They found a skew between the velocity fields of the inner region and that of the outer arms, and fitted a rotation curve which indicates that the major axis of the galaxy goes from NE to SW along PA $68 \degree$, with the NE gas receding and the SW gas approaching. Assuming trailing arms, \citet{cooke2000} concludes that the galaxy is rotating counter clockwise.

Using Chandra X-ray observations, \citet{fabbiano2011} reported the presence of two X-ray sources which they suggested were obscured AGNs, separated by $ \sim 130$ pc, with lower mass limits of $\sim 8 \times 10^{5}$ M$_{\odot}$ for the NE source and $\sim 10^{6}$ M$_{\odot}$ for the SW source.
More recent observations and analysis by  \citet{koss2015} found the same morphological correlations between the [OIII], X-ray and radio emission, but they conclude that the double SMBH detection is most likely spurious, resulting from the low number of X-ray counts ($<160$) at $6-7$ keV and data smoothing with a few counts per pixel on scales much smaller than the PSF. \\

NGC 3393 is a Compton thick galaxy \citep{koss2015} and has polarised broad $H_{\alpha}$ and $H_{\beta}$ emission lines \citep{kay2002,ramosalmeida2016}. A water maser emitting disk has been observed in the nuclear region using VLBI observations \citep{kondratko2008}: this water maser disk is observed edge on, with a major axis in $ PA \sim -34 \degree $, i.e.  perpendicular to the NLR axis. The kinematics of the water masers in the disk are consistent with Keplerian rotation, with an enclosed mass of $(3.1 \pm 0.2) \times 10^7 M_{\odot}$. \\

This paper is organised as follows. In Section \ref{sect:observations} we describe the observations and data reductions. In Section \ref{sect:results} we present the methods used for the analysis and the subsequent results. In Section \ref{sect:discussion} we present a discussion of the results and in Section \ref{sect:conclusions} we present our conclusions.


\section{Observations and Data Reduction}
\label{sect:observations}

\subsection{GEMINI-GMOS/IFU}
The observations were obtained with the Integral Field Unit of the Gemini Multi Object Spectrograph (GMOS-IFU; \citet{allington-smith2002,hook2004}) at the Gemini South Telescope on June 20, 2015 (project GS-2015A-Q-12). The observations were made in "single-slit" mode, using the IFU-R mask, and the B600+G5323 grating with four exposures of 720 s each, adding small spatial ($0\farcs5$) and spectral (50 \AA) offsets for every exposure. The spectral coverage of the observations was  $\lambda4092 - \lambda7336 $\AA, at a spectral resolution of $R = 1688$, covering the emission lines H$_{\beta}$, [OIII]$\lambda4959,5007$, [OI]$\lambda6300$, H$_{\alpha}$, [NII]$\lambda6548,6583$, and [SII]$\lambda6717,6731$, in addition to several stellar absorption lines. The standard star used for flux calibration is LTT 7987, which was observed on May 30, 2015.
The field of view of the observations was $3 \farcs 8 \times 4 \farcs 9$, which corresponds to a size of $0.96$ kpc $\times$ $ 1.24 $kpc in the galaxy, sampled at $0 \farcs 08$. \\
Seeing during the observations was $0 \farcs 62$ as measured from the FWHM of the spatial profile of the stars in the acquisition image; at the galaxy this corresponds to $155$ pc. 

The data processing was performed using tasks from the GEMINI.GMOS package for IRAF, following \citet{lena2014}. This process includes bias subtraction, flat fielding, sky subtraction, wavelength calibration, flux calibration, differential atmospheric dispersion, and finally, the building of the data cubes with a sampling of $0\farcs2 \times 0\farcs2$. The four individual data cubes are combined to avoid the detector gaps, obtaining the final data cube used throughout this paper. \\

\begin{figure*}
\centering
 \includegraphics[width=1\textwidth]{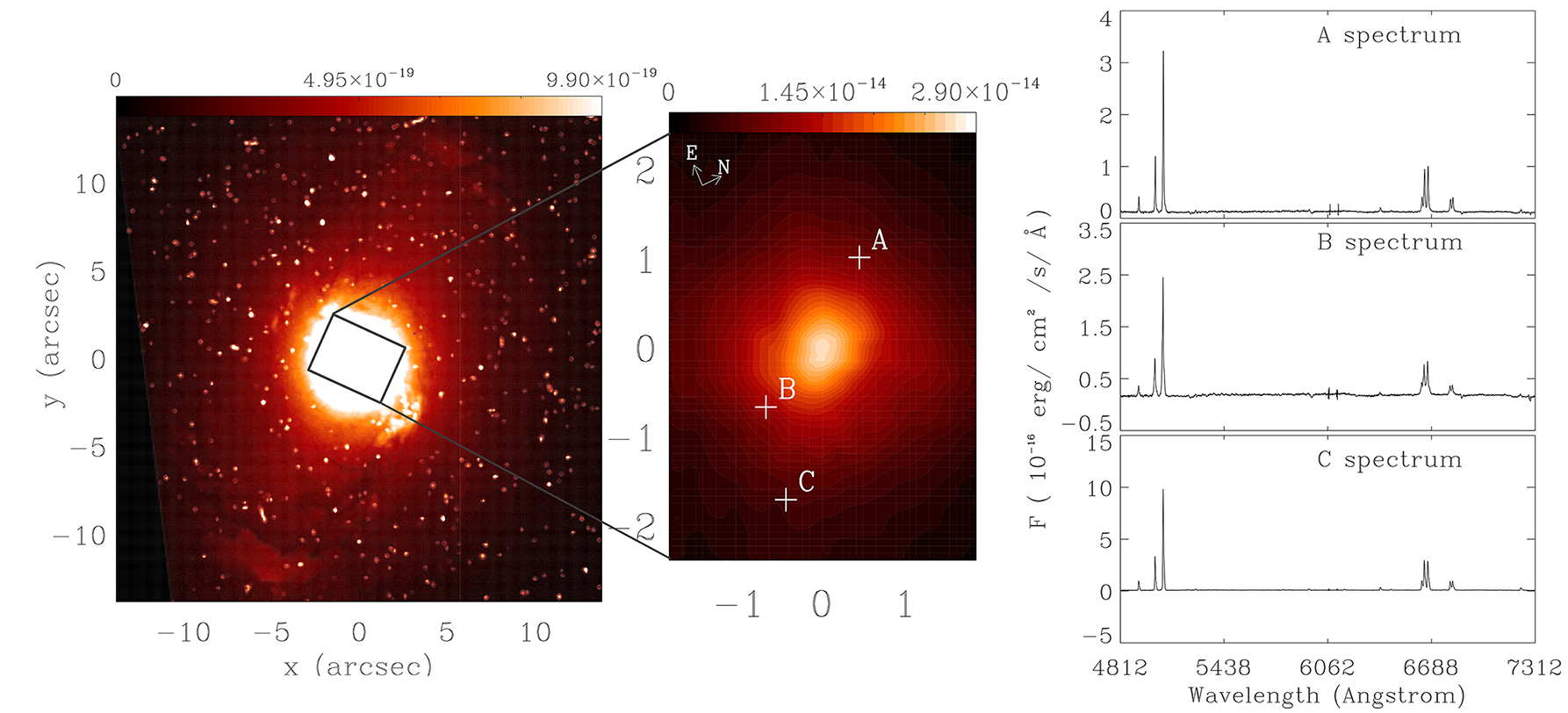}
  \caption{Left: HST F606W image for the galaxy, overlaid with the GMOS FOV. Orientation is N to the top and E to the left; Middle: GMOS continuum image, made by collapsing channels of the data cube that did not include strong emission lines. Orientation is shown with the compass in the top left corner; Right: Example spectra for the three points marked on the continuum map, showing the most prominent emission lines [OIII], [NII], H$_{\alpha}$, and [SII]. Units for both colour bars are erg/cm$^{2}$/s/\AA.}
\label{hst_and_gmoscont}
\end{figure*}
\FloatBarrier


\begin{figure}
\centering
 \includegraphics[width=0.5\textwidth]{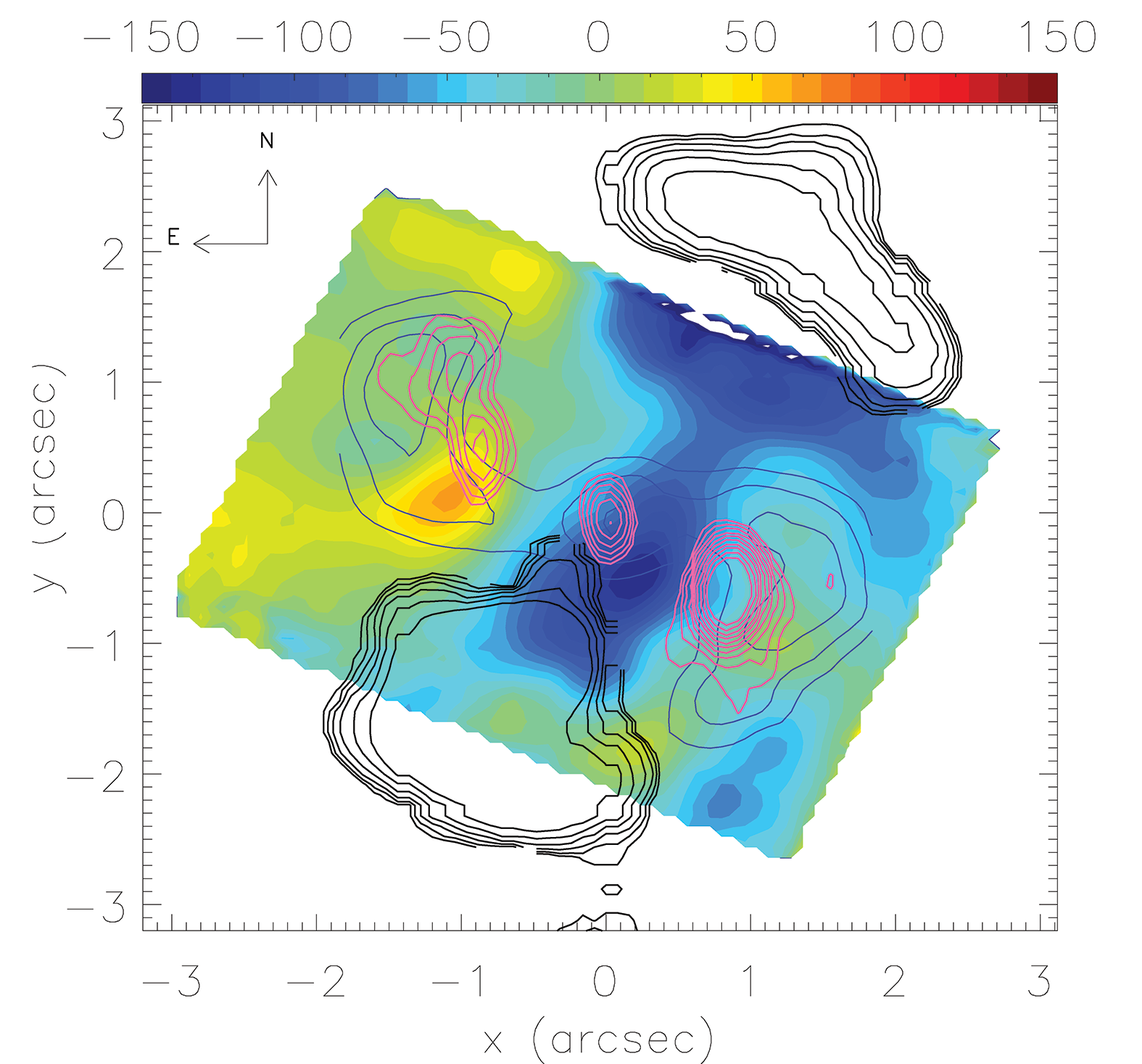}
  \caption{ The moment 1 (velocity) map of the [NII] line from our GEMINI-GMOS/IFU observations is shown in color following the color bar at the top of the panel (scale in km/s). Blue contours indicate the moment 0 map of [NII], pink contours correspond to the VLA 8.4 GHz  continuum map, and black contours show the moment 0 map from ALMA CO J:2-1 observations.
  }
\label{gmos_orientation}
\end{figure}
\FloatBarrier


\subsection{ALMA: CO J:2-1}
NGC 3393 was observed on May 3, 2016 as part of Project 2015.1.00086.S (P.I. Nagar).
Four basebands (spw's) were used: each set to an effective bandwidth of 1.875 GHz. 
Spw1 was centred on the CO J:2-1 line ($\nu_{rest} = 230.538$ GHz), with a channel
width of $2.5$ km/s. 
Spw2 was set to 'TDM' mode, for highest sensitivity, and used to cover the continuum centred on
$\nu_{rest} = 232.538$ GHz with 40.8 km/s channels. 
Spw3 was centred on the CS J:5-4 line ($\nu_{rest} = 244.935$ GHz), with a channel
width of 5.1 km/s. 
Spw4 was set to 'TDM' mode, for highest sensitivity, and used to cover the continuum centred on
$\nu_{rest} = 246.936$ GHz with 31.3 km/s channels. 
Forty one antennas were used and the total integration time on NGC 3393 was $\sim$26 min.  Six minute
scans on NGC 3393 were interleaved with one minute scans  on the nearby 'phase-calibrator'
J1037-2934. The latter is a well-studied compact quasar at redshift 0.312 with a position 
accurate to better than 1 mas. No flux calibrator was observed within this 'scheduling block'.

Data were calibrated and imaged using CASA version 4.7, and mostly followed the calibration
script provided by the ALMA observatory (the CASA calibration pipeline was not available at the time
of the release of this dataset). Since a flux calibrator was not observed for this project,
flux calibration was performed by setting the flux of the phase calibrator J1037-2934 to  
602 mJy at 235.7 GHz (a value provided in the ALMA observatories calibration script). 
The ALMA calibrator database shows that this source had a measured flux of
630 mJy when observed 12 days later at the same frequency. The continuum was imaged from line-free 
channels in all four spws.
This continuum image, at an effective frequency of 239 GHz, was made using 'Briggs weighting' with 
robust=2, i.e. 'natural' weighting. 
The synthesised beam of this image has a major (minor) axis of 0\farcs71 (0\farcs61) in
PA 85\arcdeg\  and the r.m.s. noise is 0.023 mJy/beam. 
The continuum-subtracted $uv$-data were then used to image the CO J:2-1 line.
The final CO J:2-1 data we use and show in this work come from two datacubes: 
(a) a higher spatial and spectral resolution cube, made using 'Briggs weighting' with robust=0.2,
    and using the intrinsic spectral resolution. This datacube has a synthesised beam with
    major (minor axis) of
    0\farcs58 (0\farcs5) with a beam PA of $-$72\arcdeg\ and an r.m.s. noise of 
    0.7 mJy/beam per 2.5 km/s channel;
(b) a lower spatial and spectral resolution (but higher signal to noise) cube, made using
   'natural' weighting, and 4-channel spectral averaging. This datacube has
     a synthesised beam with major (minor) axis of 
     0\farcs73 (0\farcs62) with a beam PA of 86\arcdeg\ and an r.m.s. noise of 
     0.45 mJy/beam per 10 km/s channel. \\

 \section{Results}
 \label{sect:results}

\subsection{Stellar kinematics}
\label{subsect:stellar_kinematics}

The stellar kinematics was obtained from the absorption lines in the GEMINI-GMOS/IFU datacube. To model the stellar kinematics we used the penalised pixel-fitting (pPXF v5.2.1) routine developed by \citet{cappellari2004} and upgraded in \citet{cappellari2017}, where the line-of-sight velocity distribution (LOSVD) is recovered by fitting an optimised template to the galaxy spectrum. We used the INDO-US spectral templates library \citep{valdes2004}.
To reach the highest possible S/N to measure the stellar kinematics reliably, we spatially binned the data cube (S/N $=50$) by using the Voronoi binning method described in \citet{cappellari2003}.  The spectra show no absorption lines in the region 
near the northern radio lobe: this region was thus masked  before running pPXF. \\

\begin{figure}
\centering
\includegraphics[width=.5\textwidth]{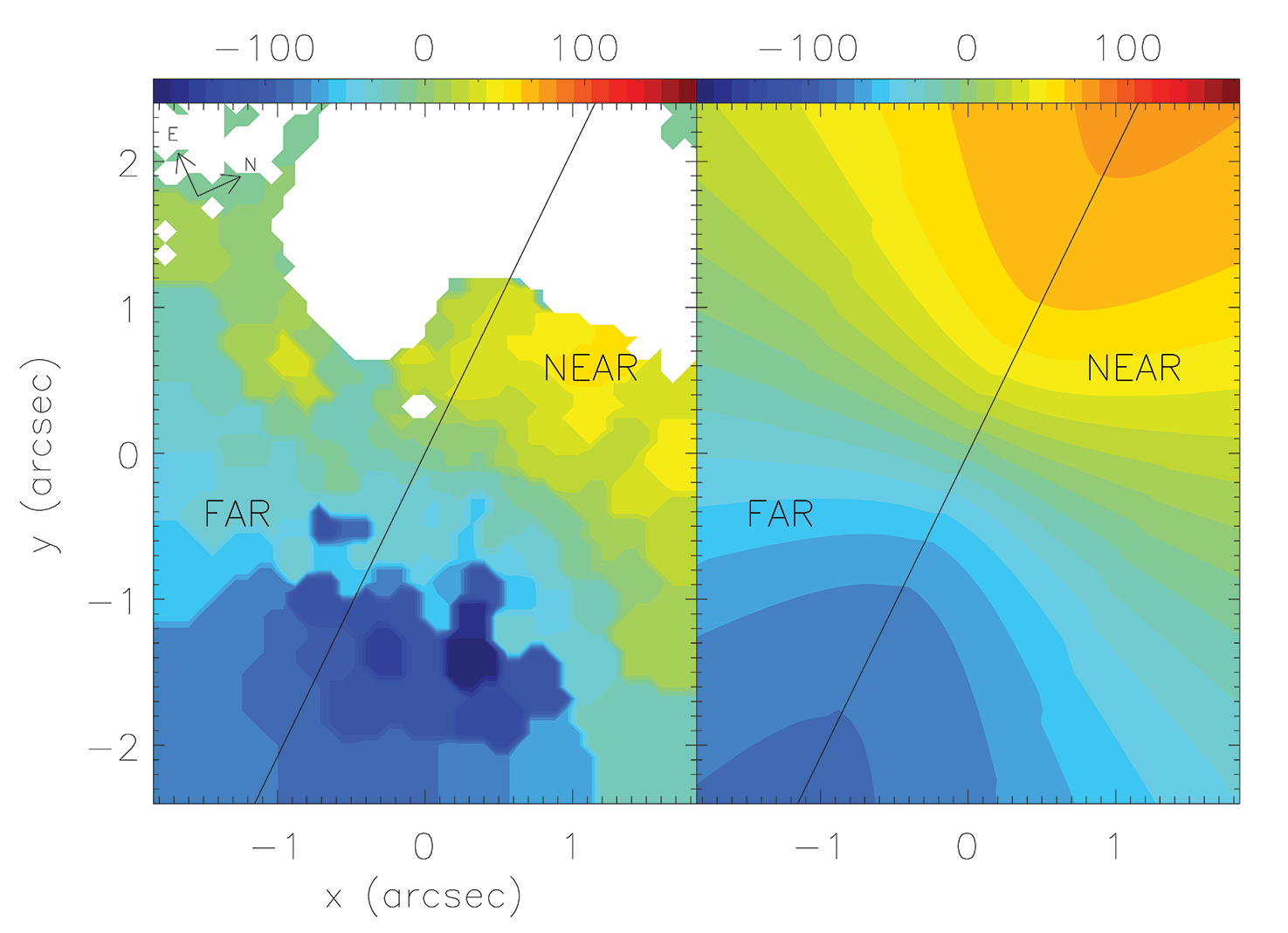}
\label{stellar_velmap}

\caption{ Left: Stellar velocity map from pPXF; regions where no absorption line information could be recovered are masked. Right: best-fit Bertola model to the stellar velocity field; centre and inclination ($i=25 \degree$) were kept fixed. The free parameters fitted were $A=207$ km/s, $C=0.82$, $p=1$, and $PA = 40 \degree$.
}
\label{stellar_kinematics}
\end{figure}

The pPXF-derived stellar kinematic velocity map is shown in Fig. \ref{stellar_kinematics}. Throughout this paper we adopt an inclination of $25\degree$ based on the axis ratio $a/b = 1.1$ \citep[v1.10,][]{devaucouleurs1991}. We model the stellar velocity field obtained from pPXF using a spherical potential with pure circular rotation, assuming that the kinematical centre is cospatial with the peak in the continuum emission. The observed radial velocity from this potential is given by \citep{bertola1991}:

$$ V = V_{sys} + \frac{AR\cos(\psi - \psi_{0})\cos^{p}\theta }{ (R^{2}[\sin^{2}(\psi - \psi_{0}) + \cos^{2}(\psi - \psi_{0})]  + c^{2}\cos^{2}\theta )^{p/2} } \; , $$

\noindent where $V_{sys}$ is the systemic velocity, $R$ is the radius, $\theta$ is the disk inclination, $\psi_{0}$ is the position angle of the line of nodes, $A$ is the amplitude of the rotation curve, $c$ is the concentration parameter regulating the compactness of the region with a strong velocity gradient, and $p$ regulates the slope of the 'flat' portion of the velocity curve.
We perform a least-square minimization using the IDL routine MPFIT2DFUN \citep{mpfit-markwardt2009} to obtain the best fitting parameters. The resulting model is shown in Fig. \ref{stellar_kinematics}, where we kept centre and inclination ($25\degree$) fixed, and the free parameters of the fit and the best fitted values were: $A=207$ km/s, $c=0.82$, $p=1$, $\Psi_0 = 40 \degree$.

\subsection{Ionised gas}
\label{sect:ionised_gas}

To model the ionised gas emission in the GMOS-IFU data we use custom IDL routines. We begin our analysis by generating moment images for the most prominent spectral emission lines, [OIII]$\lambda 5007$ and the [NII]$\lambda6549,\lambda6585$ doublet. These moment images are created by collapsing one axis of the data cube.

\begin{figure*}
\centering
\includegraphics[width=0.99\textwidth]{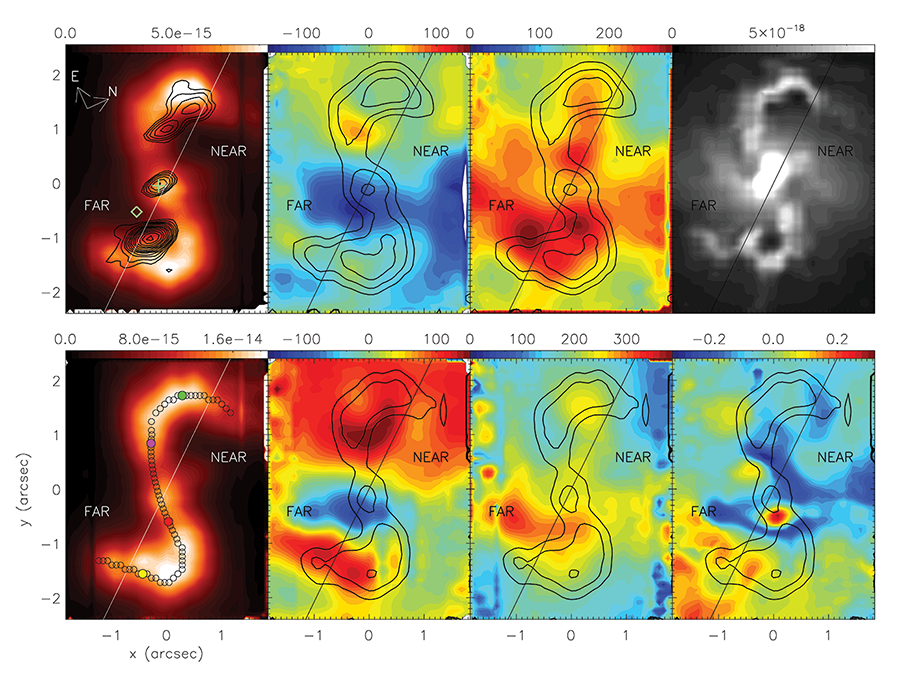}

\caption{Moment maps for the [NII]$\lambda6585$ (top) and [OIII]$\lambda5007$ (bottom) emission lines. First column: integrated flux, second: velocity map, third: velocity dispersion. The fourth column shows the structure map (top) and the h3 moment (bottom) from a one-component Gaussian-hermite fit to the [OIII] emission line; this moment represents the asymmetric deviations from a Gaussian profile.
Black contours superposed on the moment 0 map of the [NII] line correspond to the VLA 8.4 GHz continuum map. Black contours in the moment 1, 2 and 3 maps correspond to the moment 0 of the respective line. Black circles in the moment 0 map of the [OIII] line (bottom left corner) show the apertures positions along an s-shaped slit used to extract the position-velocity diagram shown in Fig. \ref{gmos_pvdiagram}, the coloured apertures mark specific regions that can be identified in the pv-diagram.
The galaxy major axis (PA $40 \degree$) is marked as a black line, delimiting the near and far side. The compass shown on top left corner shows the orientation of our GEMINI-GMOS/IFU data. The green cross marks the position of the stellar continuum peak, which we assume to trace the position of the nucleus. The green diamond shows the position of the SW secondary BH reported by \citet{fabbiano2011}.
Colour bar units for moment 0 maps and structure map are erg/cm$^{2}$/s/\AA  for, km/s for moments 1 and 2, and unitless for h3 moment.
}
\label{onecomp_gh_gmos}
\end{figure*}

The moment zero, i.e., integrated flux maps (Fig. \ref{onecomp_gh_gmos}) show an S-shaped morphology of the ionised gas, where two arm-like features leave the centre as a straight line along PA $55 \degree$ and then curve, to the NW in the NE arm, and to the SE in the SW arm. The brightest emission is observed within an opening angle of $90 \degree$ from the nucleus.
The black contours in the [NII] moment zero map (Fig. \ref{onecomp_gh_gmos}) correspond to the 8.4 GHz VLA map, and indicate that the gas in the S-shaped arms seems to surround both NE and SW radio lobes.
This morphology, and the interaction between radio jet and ionised gas, was previously observed and analysed by \citet{cooke2000}, and \citet{maksym2017}. \\

Moment one, i.e., velocity maps, are shown in the second column of Fig. \ref{onecomp_gh_gmos}, where a gradient can be observed from NE to SW. However, given the complex kinematics of the NLR a precise determination of its PA is not possible using this moment image. Two high-velocity features are found to the NE and SW of the nucleus. The moment one map for the [OIII] emission line shows that a redshifted component covers a large fraction of the FOV. For the [NII] line we observe that the NE region shows larger redshifts with increasing distance from the nucleus. However, with the exception of the blueshifted blob observed S of the nucleus, the SW region is not as blueshifted as expected from an inclined disk in pure rotation. \\

The moment two (velocity dispersion) maps, shown in the third column of Fig. \ref{onecomp_gh_gmos}, presents a large dispersion in the central region, extending from the centre in a section along the minor axis (referred to as the equatorial region hereafter), and a second high dispersion area is seen in the inner part of the NE arm. The dispersion is higher for the [OIII] emission line. \\

The moment 3 (h3) map (bottom-right corner in Fig. \ref{onecomp_gh_gmos}) obtained from a one-component Gaussian fit, describes asymmetric deviations from a Gaussian profile, presents some skewness in the equatorial region, where negative values are indicative of a blueshifted wing or component. A large area in the SE region present a positive skewness, which indicates a strong redshifted wing or component. 
Similar distributions were observed for all moment maps of all strong emission lines fitted with a single Gaussian using PROFIT ([SII], H$\alpha$, H$\beta$, [OI]; not shown). \\

\begin{figure}
\centering
\subfloat{
\includegraphics[width=.5\textwidth]{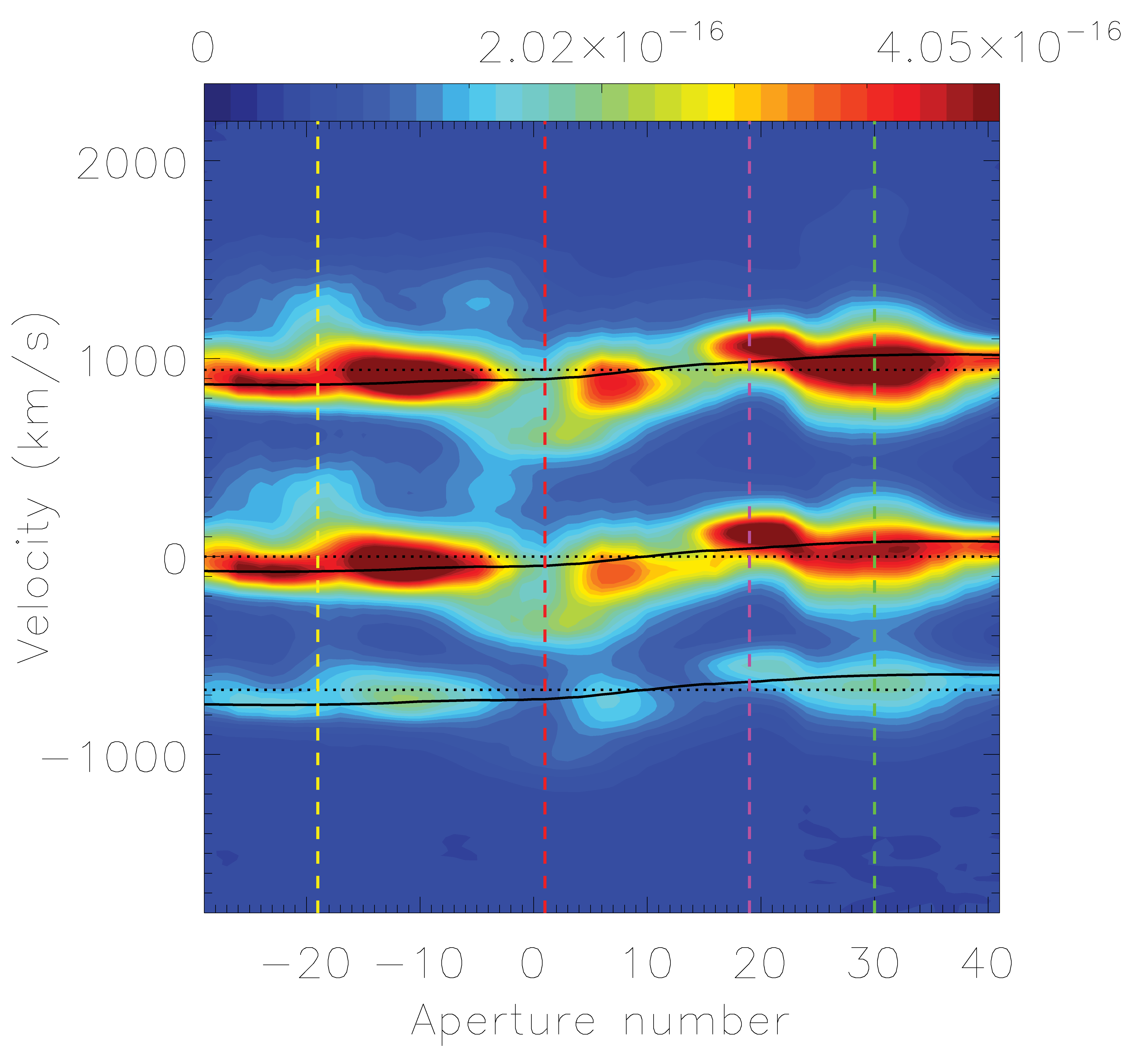}
\label{gmos_pvdiagram_ha}
}
\caption{Position-velocity diagram of the continuum-subtracted GEMINI-GMOS/IFU data cube, centred on the H$_{\alpha}$ emission line, extracted along the S-shaped 'slit' shown on bottom right panel of Fig. \ref{onecomp_gh_gmos}. The solid black line shows the expectations of the Bertola rotation model derived from the stellar kinematics. For reference, vertical yellow, red, magenta and green lines show the position of specific apertures that are marked with the same colours in Fig. \ref{onecomp_gh_gmos} (bottom left corner). The dashed black lines show the zero-velocity for each emission line. Colour bar units are ergs/cm$^{2}$/s/\AA.
}
\label{gmos_pvdiagram}
\end{figure}

To better examine the kinematics we defined a curved 'slit' which closely follows the S-shaped arms seen in the emission lines: the aperture positions of this slit are shown in Fig. \ref{onecomp_gh_gmos} (bottom left panel) and the position-velocity (pv) diagrams of the $H\alpha$ and [NII] lines, extracted from these apertures are shown in Fig.\ref{gmos_pvdiagram}. The pure rotation model fitted to the stellar kinematics is shown as the solid black line. While we do see some gas at velocities close to this model, there are large deviations from the model which indicate the presence of multiple kinematic components.  To guide the eye specific apertures along the slit are coloured (Fig. \ref{onecomp_gh_gmos}), and the pv-diagram is marked with a corresponding colour line at this aperture.  The yellow and red apertures are near the SW radio lobe, while the magenta and green line apertures are near the NE radio lobe.  The apertures near the yellow vertical line reveal gas that appears to follow the pure rotation model plus a redshifted wing with velocities reaching 500 km/s; no equivalent blueshifted wing is observed. In the apertures near the red vertical line we see little to no gas following the rotation model: instead we observe strongly blueshifted emission with velocities near $- 400$ km/s. Apertures near the magenta line show a small fraction of gas following the rotation model, and a dominant component of gas is redshifted by up to $\sim 300$ km/s. Apertures near the green line show a highly broadened profile: while the median velocity is roughly close to the rotation model, we see large ($\pm$300 km/s) redshifted and blueshifted velocities. In this region, an additional extreme redshifted component is observed, most clearly seen in the [NII]$\lambda$6585 line: this weak redshifted wing reaches velocities of 1000 km/s. 
Similar characteristics are observed in the pv-diagram of the [OIII] emission line (not shown).

The kinematics of NGC 3393 were classified as turbulent by \citet{fischer2013}, based on HST spectroscopy, since they could not be satisfactorily fitted with a biconical outflow model.
Given the complex kinematics of the NLR, the large velocity dispersion, the Gaussian skewness observed in the equatorial region, the multiple kinematical components observed in the pv-diagram, and a visual inspection of the spectra in the mentioned areas (Fig. \ref{multicomp_spectra}), the need for a multiple-component Gaussian fit is clear. \\

\begin{figure*}
\centering
\includegraphics[width=0.79\textwidth]{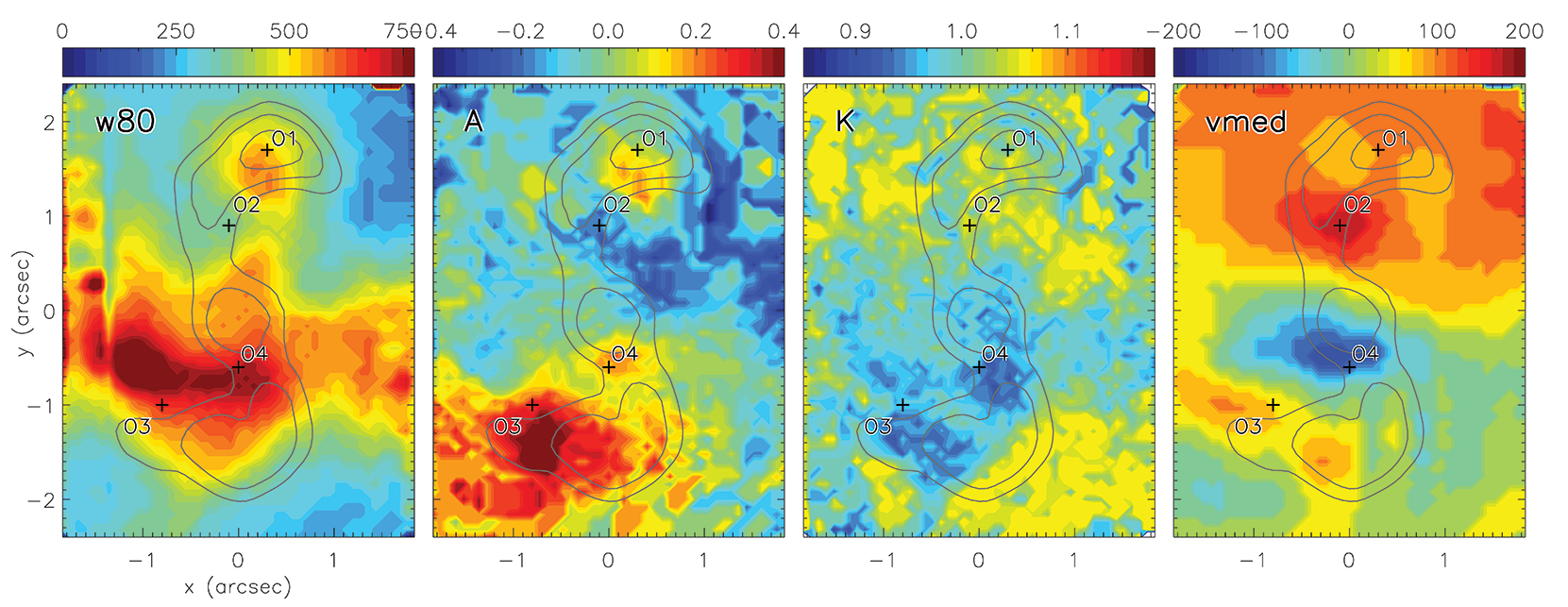}

\caption{Profile measurements for emission line profiles described in \citet{liu2013,whittle1985}. From left to right: W80 (velocity width that encloses 80 per cent of the total flux), A (asymmetry of the profile parameter), K (shape of the profile parameter) and V$_{med}$ (median value of the integrated flux profile). For W80 and V$_{med}$ the colour bar is in km/s, while A and K are unitless parameters. Grey contours correspond to the [NII] moment 0 map. Labels 'O1', 'O2, 'O3' and 'O4' correspond to areas of interest, explained in Sect. \ref{sect:ionised_gas}.
}
\label{lineprof_par}
\end{figure*}

A visual inspection of the emission line profiles for the [OIII], H$\alpha$ and [NII] lines show a large difference of shape and width of the profiles in different regions of the FOV. To understand this difference we use the measurements described in \citet{whittle1985}, where velocity widths are measured at some fraction of the cumulative line flux, the integral nature of these measurements makes them relatively insensitive to the nature of the profiles. For every spaxel we calculated: (a)\textbf{W80}, a line width parameter which measures the velocity width that encloses 80 per cent of the total flux. This parameter does not discard information of broad wings. (b) \textbf{A}, an asymmetry parameter as defined in \citet{liu2013}, where a symmetric profile will have a value of A$=0$, and the presence of redshifted (blueshifted) wings will give a positive (negative) value. (c) \textbf{K}, a shape parameter that is related to the line kurtosis, for a Gaussian profile K$=0$, while profiles that have broad wings will have K $>1$ and stubby profiles will have K $<1$. The values obtained for these parameters for the [OIII] emission line are shown in Fig. \ref{lineprof_par}. We use these results to identify spaxels in the FOV for which multiple component fits are required and factible. We consider pixels for which  W80 $> 350$ km/s, A $> 0.1$ or A $<-0.1$,  and K $>1.1$ or K $< 0.8$, we created a mask by weighting the values for W80, A and K, to 80, 20 and 20 per cent respectively. furthermore, regions with low S/N are masked out as the profiles are not as reliable. This mask contains the spaxels that require two Gaussian fits. Visual inspection of the maps show considerable larger values for W80 in some areas. For this area we consider the possibility of a three Gaussian component fit, and thus we made a second mask for regions where W80 $>$ 500 km/s.

\begin{figure*}
\centering
\includegraphics[width=0.72\textwidth]{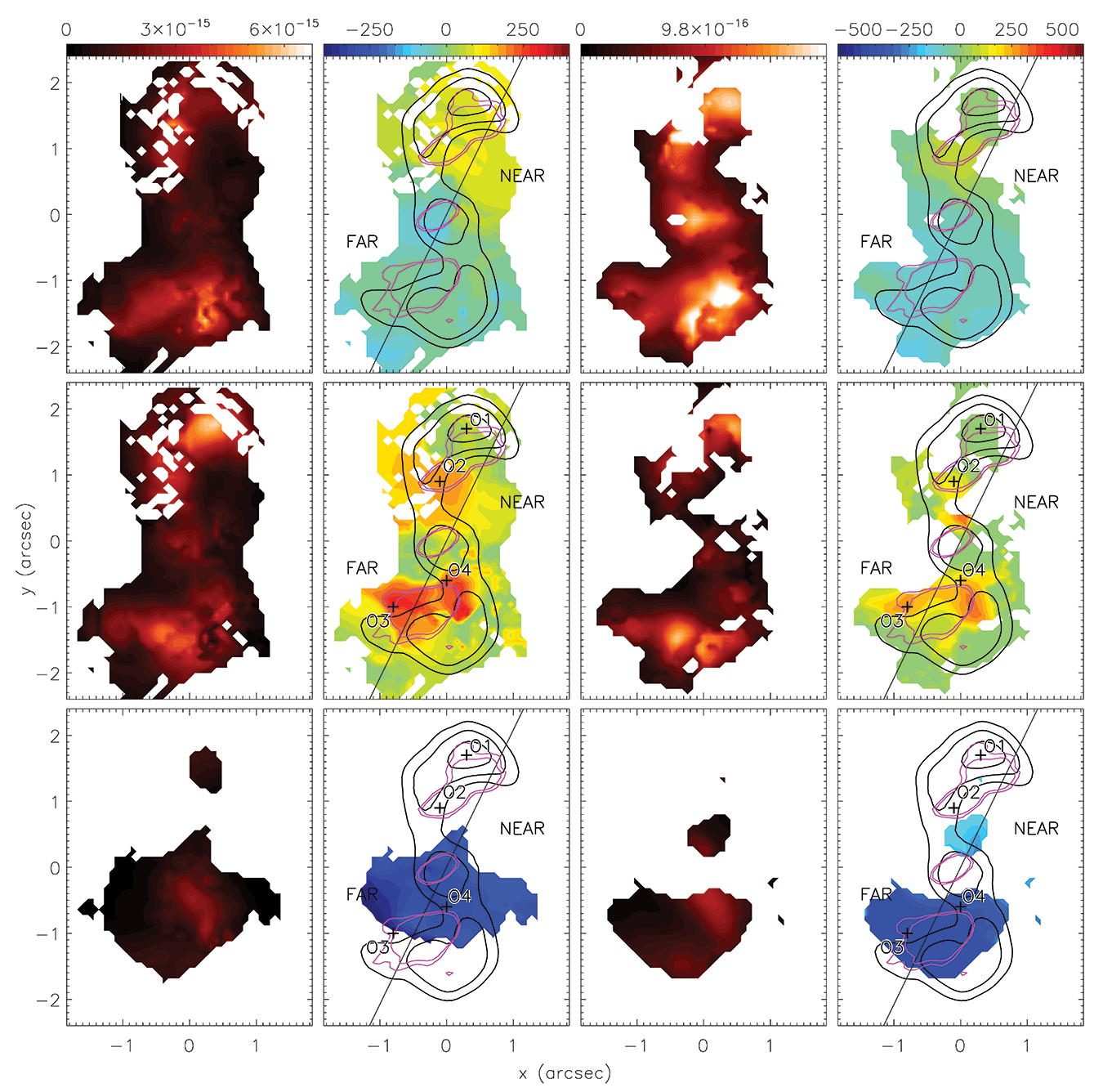}

\caption{Moment 0 and 1 maps from our multiple-component Gaussian fits to the [OIII] and [NII] emission lines. 
First and second column show the moment 0 and moment 1 map for the [OIII] line. Third and fourth column show the moment 0 and 1 for the [NII] line. First, second and third row show the narrow, broad redshifted and broad blueshifted components respectively. Crosses marked as 'O1', 'O2', 'O3', and 'O4' show the areas of interest defined in the text. Black contours in the moment 1 maps show the moment 0 map of the corresponding emission line, obtained from the axis collapse, as shown in Fig. \ref{onecomp_gh_gmos}. Pink contours correspond to the VLA 8.4 GHz map. Units for the moment 0 maps are erg/cm$^{2}$/s/\AA , and km/s for the moment 1 maps.  }
\label{gmos_multicomp_mom1}
\end{figure*}

\begin{figure}
\centering
\includegraphics[width=0.5\textwidth]{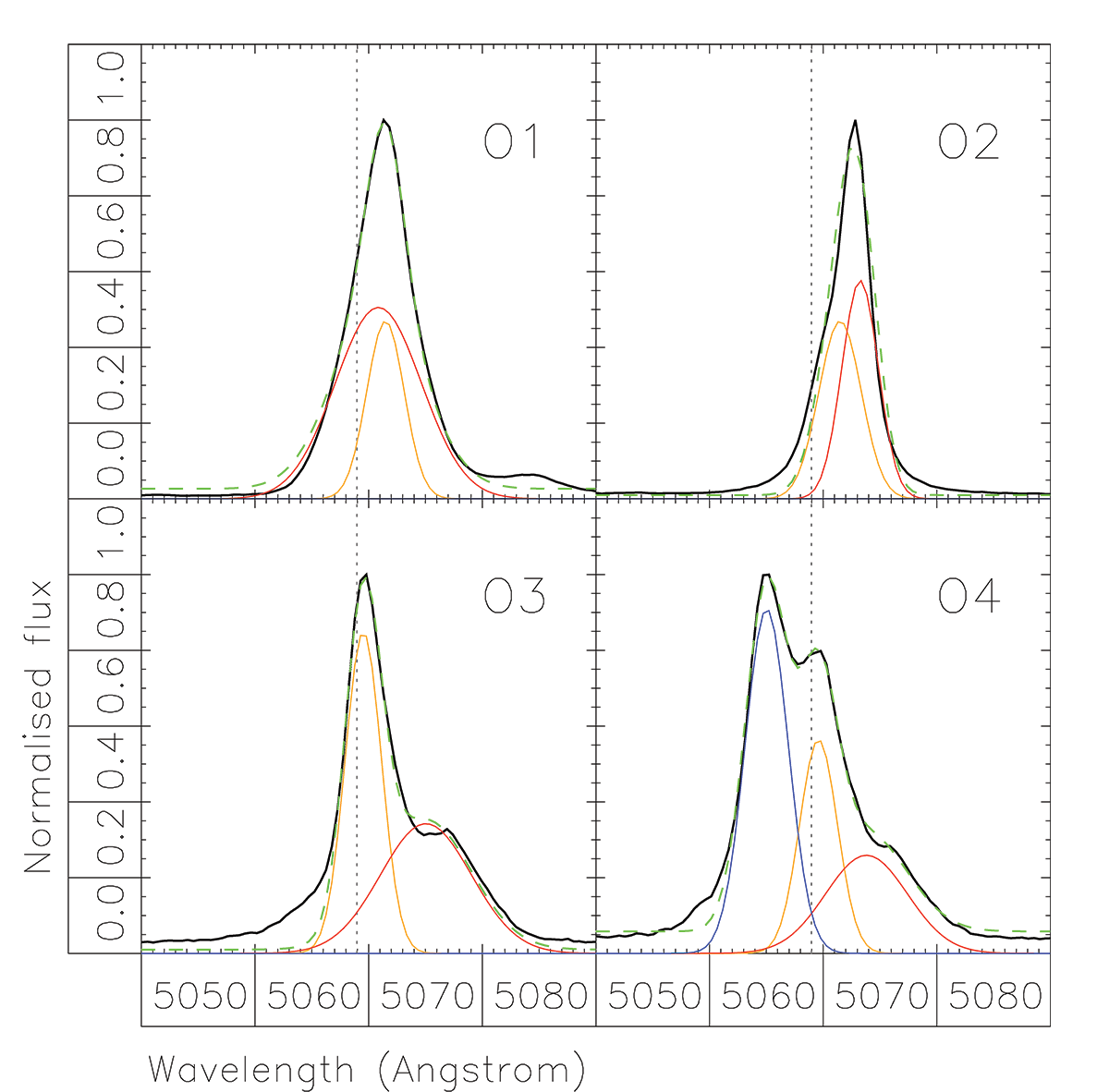}

\caption{ Examples of multiple-component Gaussian fits to the [OIII] emission line in the 'O1', 'O2', 'O3', and 'O4' areas shown in Fig. \ref{gmos_multicomp_mom1}. The narrow, broad redshifted, and broad blueshifted components are shown in yellow, red, and blue, respectively and their sum is shown by the dashed green line. Dotted vertical line shows systemic velocity.
}
\label{multicomp_spectra}
\end{figure}

For the multi-component analysis we use a custom IDL routine based on the routine PROFIT \citep{riffel2010}, fitting multiple Gaussian profiles to the emission lines of the GMOS-IFU spectra. We performed a two and three component Gaussian fit to the emission lines. We use  a narrow component ($ \sigma < 115$ km/s ), a broad redshifted component ($ 115 < \sigma < 230$ km/s), and a third broad blueshifted ($ 115 < \sigma < 230$ km/s) component,  based on the masks described above. Note that the 'broad' nomenclature used here should not be confused with the broad component typical of a Type 1 AGN.\\
The result of the multiple-component fit is shown in Fig. \ref{gmos_multicomp_mom1}. To fit the narrow component we use as a first guess for each spaxel the velocity predicted by the pure rotation model fitted to the stellar kinematics, this component traces, more or less, gas rotating in a disk, with a kinematic major axis in PA $40 \degree$ (this value is in agreement with the $37 \pm 3\degree$ obtained by \citet{cooke2000}). From the moment zero maps it can be observed that this component is weaker for the [OIII] line which seems to be more dominated by the broad red and blueshifted components.
The broad redshifted components trace more complex kinematics which seem to be closely related to the radio jets. We have marked four areas of more interesting kinematics  (regions marked 'O1', 'O2', 'O3' and 'O4' labels in Fig. \ref{gmos_multicomp_mom1}), these areas coincide with the coloured vertical lines we used in the pv-diagram. The 'O1' area, as we observed in the pv-diagram, has a broad profile (top left panel on Fig. \ref{multicomp_spectra}) that is slightly asymmetric, however no clear multiple peaks are observed and the separation between model and disturbed gas is not clear. A very broad weak redshifted wing is also observed in this area. The regions 'O2' and 'O3' seem to be outflow dominated and have strong high-velocity redshifted emission and appear to be closely related to the NE and SW radio lobes respectively. 
The 'O4' area shows the widest profiles observed in the FOV (bottom right panel in Fig. \ref{multicomp_spectra}), where three clear peaks are observed, the dominant component is the blueshifted emission, followed by a component that seems to follow rotation on the disk, and a weaker redshifted wing. This profile extends along a region perpendicular to the radio jet axis.

\subsection{Molecular gas : CO J:2-1}
\label{subsect:molecular_gas}

\begin{figure}
\centering	
  \includegraphics[width=0.49\textwidth]{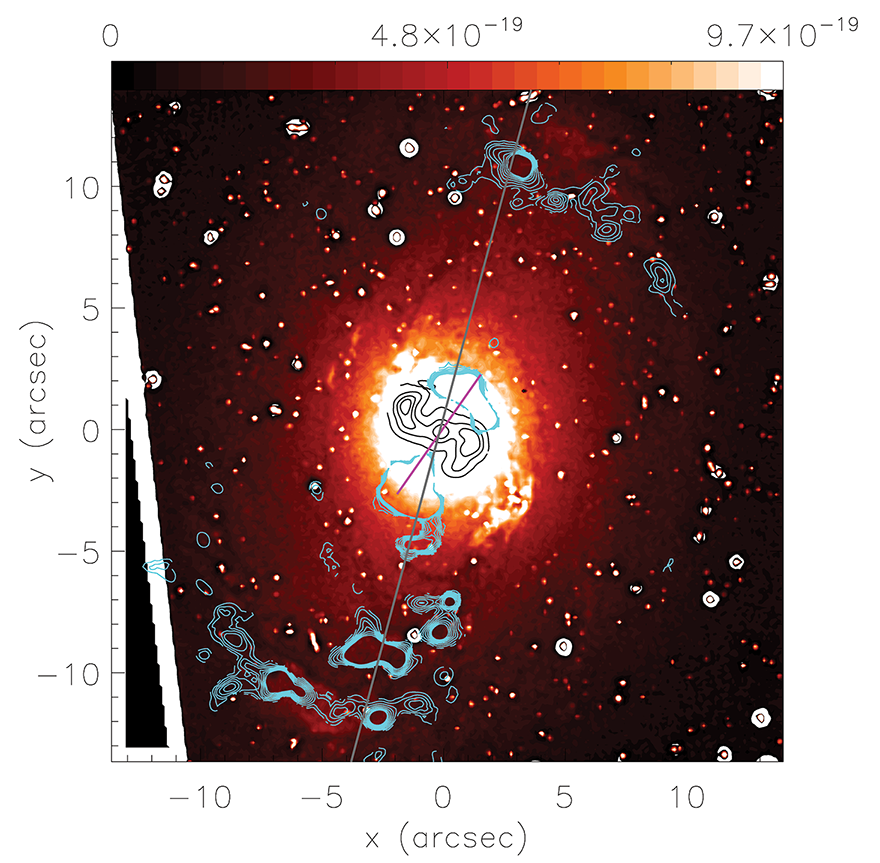}
  \caption{The structure map (obtained using a F606W HST image) of NGC 3393 is shown in colour, with overlays of the ALMA CO J:2-1 moment 0 map (cyan contours) and the GMOS [NII] moment 0 map (black contours). The grey line marks the PA of the large-scale bar, and the magenta line corresponds to the PA and estimated extension of the nuclear bar. Colour bar units are ergs/cm$^{2}$/s/\AA.
  }
\label{overlay_hst}
\end{figure}

\begin{figure}
\centering
\includegraphics[width=0.5\textwidth]{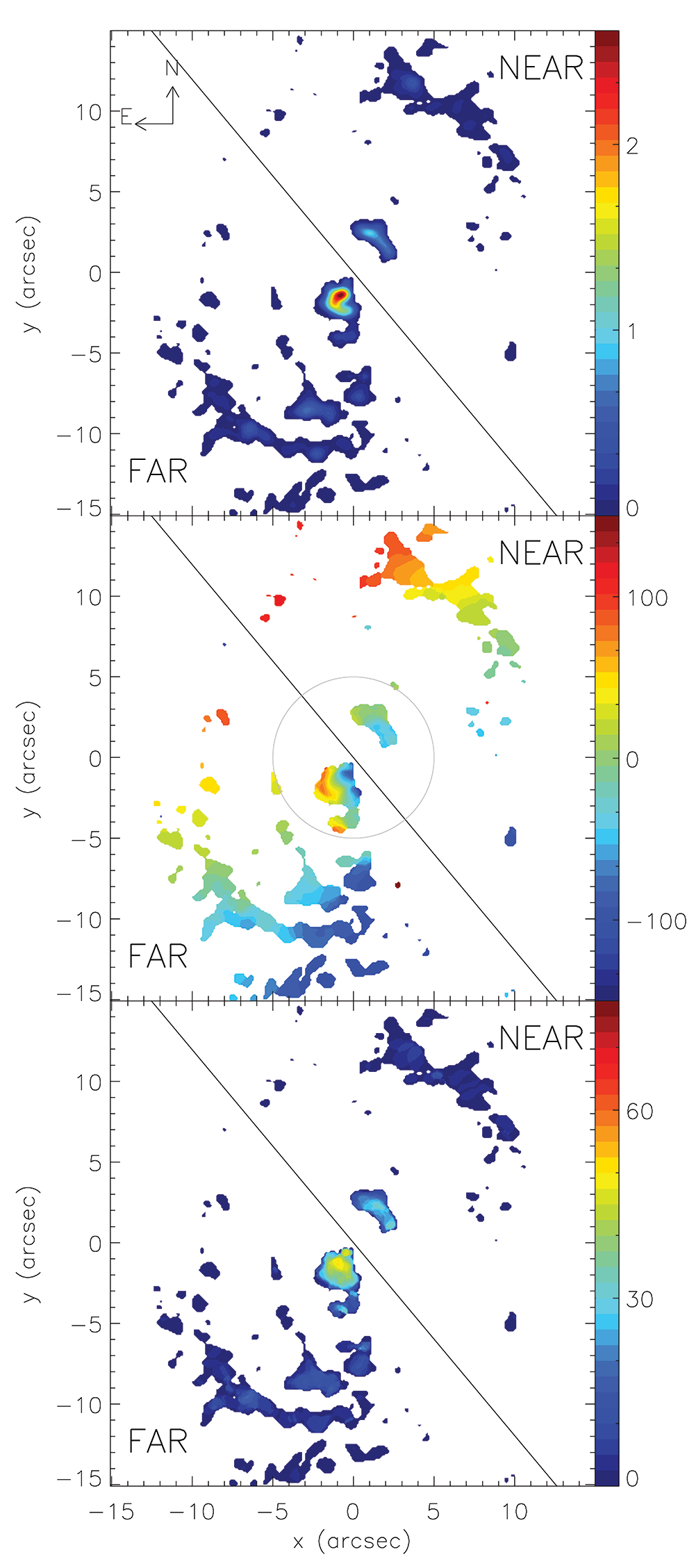}

\caption{Moment maps for ALMA CO J:2-1 data. Top: integrated flux (moment 0) map following the colour bar (units of Jy/beam). Middle: velocity map (moment 1) after subtraction of a CO systemic velocity of $3746$ km/s; the grey circle separates the inner region and outer region referred to in the text. Bottom: velocity dispersion (moment 2) map. Moment 1 and 2 colour bar units are km/s. In all panels N is up and E is to the left, and the black line marks the adopted major axis PA of $40 \degree$.
}
\label{alma_moments}
\end{figure}

To illustrate the distribution of the molecular gas relative to the other components of the galaxy we show in Fig. \ref{overlay_hst} both the CO J:2-1 (ALMA; cyan) and [NII] (GEMINI-GMOS/IFU; black) integrated flux maps overlaid on an HST F606W image to which we applied an unsharp-mask filter, with the goal of emphasising structures, such as dusty regions (we refer to this filtered image as 'structure map' from now on). It can be observed that the ionised and molecular gas barely overlap, due to the limited FOV of our GMOS-IFU data, this can be observed better in Fig. \ref{gmos_orientation}. There is little to no CO J:2-1 molecular gas observed in the central region. This implies either a true lack of molecular gas or is a critical density effect, that is, the gas density may be high enough that the CO J:2-1 transition is collisionaly, rather than radiatively, de-excited. The outer distribution of CO J:2-1 seems to follow the inner ring-like structure, where the structure map shows the presence of dust.
The observed molecular gas morphology in the nuclear region could be a result of the molecular gas and radio jet interaction: the latter can produce entrainment of the gas along the jet P.A, and push gas away from the centre perpendicular to this PA. Alternatively, the molecular gas density is high enough that the CO J:2-1 transition is 'dark' and a dense molecular gas tracer is required.

The CO J:2-1 moment 0 map is shown in Fig. \ref{alma_moments}: the distribution of the molecular gas is fragmented and does not cover a large fraction of the field of view. The SE component close to the centre is the region with the  brightest CO J:2-1 emission. The outer region show large scale spiral arms that broadly resemble a ring. There is some emission present in the region between the SE component and the outer arms. \\

Figure \ref{alma_moments} presents the CO J:2-1 moment 1 map. We classified the kinematics in two regions, an inner region inside a circle of $5 \arcsec$ (marked in grey in Fig. \ref{alma_moments}) and an outer region, outside this circle. We can see a gradient in velocity from NE to SW in the outer distribution of gas, as it is expected for gas rotating in the disk. However, in the inner region the kinematics does not follow the outer region's rotation: in the SE inner component there seems to be a gradient of velocity in PA $-45 \degree$, and the NW inner component shows mainly blueshifted velocities.

We model the kinematics in the outer region with a pure circular rotation model obtained from an exponential disk potential, defined by:

$$ \Phi(R,z) = -2\pi G \Sigma_{0} r_{d}^{2} \int_{0}^{\infty} \frac{J_{0}(kR)e^{-k|z|} }{[ 1+(kr_{d})^{2} ]^{3/2}} dk \; ,$$

\noindent where $(R,z)$ are cylindrical coordinates, $G$ is the gravitational constant, $\Sigma_{0}$ is the central surface brightness, $r_{d}$ is the disk scale length, and $J_{0}$ is the zeroth order cylindrical Bessel function. For this model we assume an infinitesimally thin disk ($z \rightarrow 0$). The rotation velocity from this potential is given by

 $$V^{2}_{ROT} (R) = R \frac{\partial \Phi}{\partial R} \; $$

\begin{figure}
\centering
\includegraphics[width=0.5\textwidth]{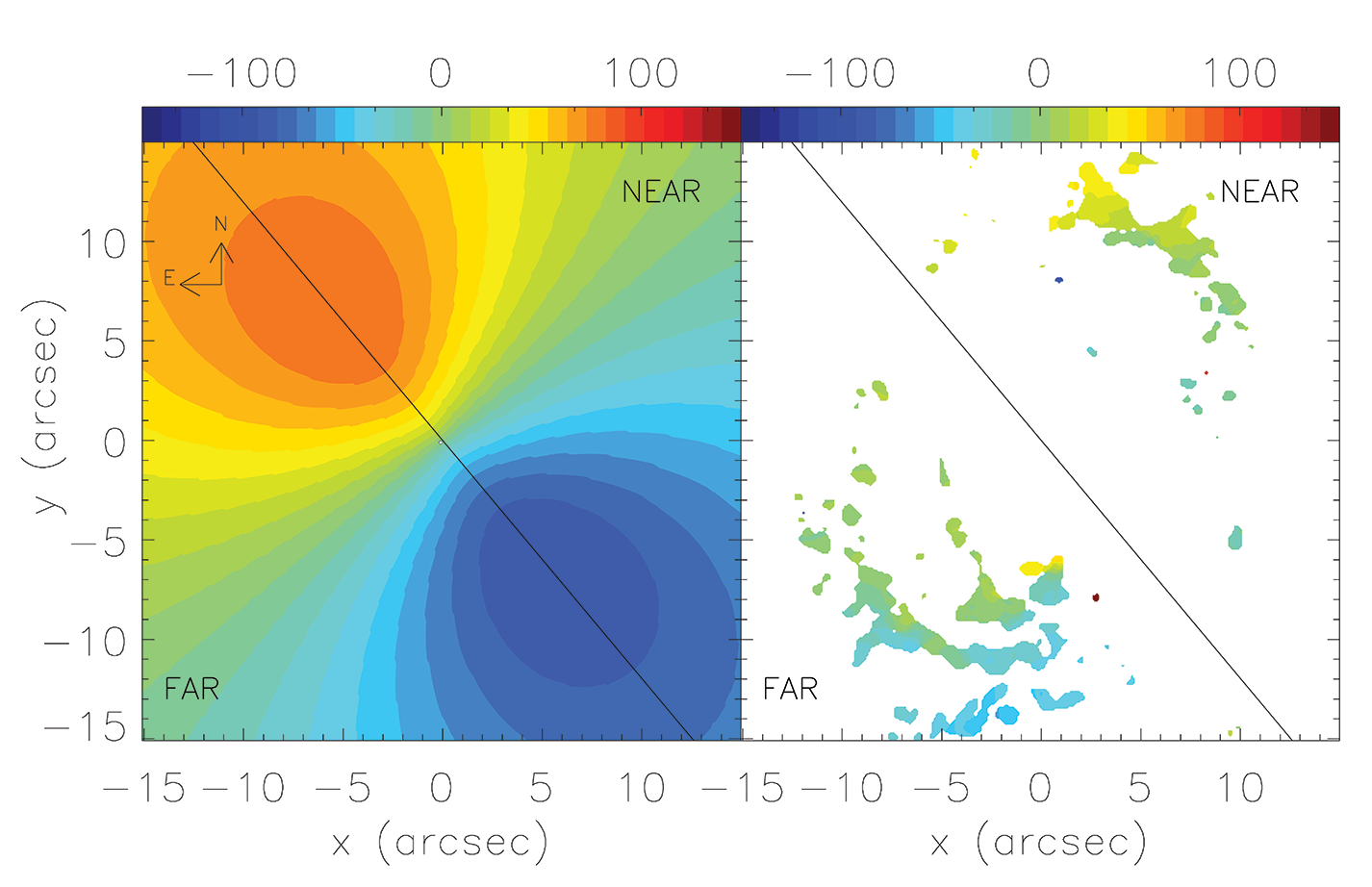}

\caption{Left: Our pure rotation model derived from fitting the outer (outside the grey circle in Fig. \ref{alma_moments}) CO velocity field with a model based on an exponential disk potential (see text). Right: residual (observed - model) velocity field of the outer disk.}
\label{alma_model_expdisk}
\end{figure}

\begin{figure}
\centering	
\includegraphics[width=0.5\textwidth]{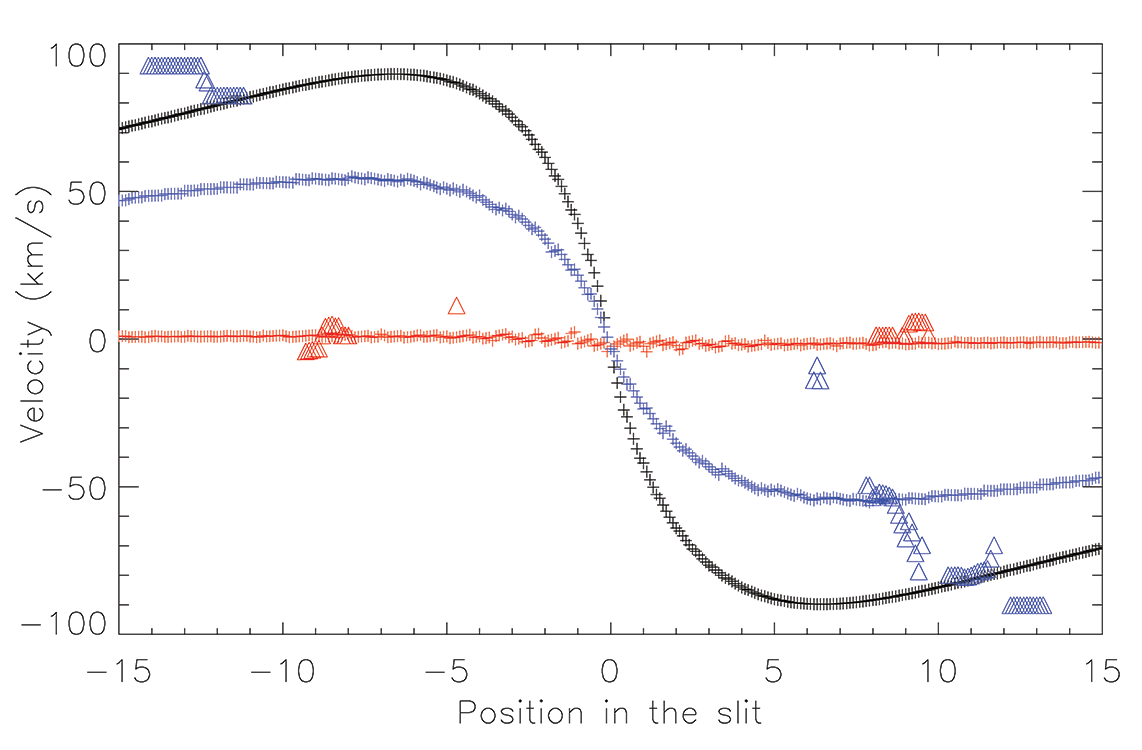}
  \caption{ CO J:2-1 rotation curves in the outer (outside the grey circle in Fig. \ref{alma_moments}) region along the major axis, minor axis (red triangles) and PA $10 \degree$ (blue triangles). Plus symbols show the best-fitted pure rotation exponential disk model along the same PAs.
 }
\label{rotcurve_alma_major_minor_axis}
\end{figure}

\noindent We perform a least-square minimization (using the MPFIT2DFUN routine in IDL) to obtain the parameters that best fit the observed CO J:2-1 velocity field (Fig. \ref{alma_model_expdisk}). The PA cannot be well constrained due to the scarcity of the data along the major axis, we thus fix the major axis PA to $40 \degree$ (See Sect. 5) after verifying that the velocity profiles of the CO J:2-1 along, and near, the minor axis is most consistent with this major axis PA (e.g., Fig. \ref{rotcurve_alma_major_minor_axis}), and also fix the inclination to $i = 25 \degree$. Based on the rotation curve along the minor axis, we use a $-26$ km/s offset from the systemic velocity \citep[$V_{sys} = 3746$ km/s from HIPASS;]{hipass-meyer2004}, to better fit the CO J:2-1 data.
For the free parameters $r_{d}$ and $M_{d}$, the mass and scale length of the disk respectively, the best fitted values obtained are $r_{d} = 1 $kpc and $M_{d} = 2.7 \times 10^{10}$ M$_{\odot}$.
For comparison, the values obtained by the HST light distribution modelling of \citet{lasker2016} are $r_{d} = 1.38$ kpc and $M_{d} = 2.04 \times 10^{10}$ M$_{d}$. The resulting exponential disk model and the residual from the model subtraction to the CO J:2-1 velocity map are shown in Fig. \ref{alma_model_expdisk}. The rotation curves of the CO J:2-1 and the exponential disk model are shown in Fig. \ref{rotcurve_alma_major_minor_axis}. \\

\begin{figure}
\centering
\includegraphics[width=0.5\textwidth]{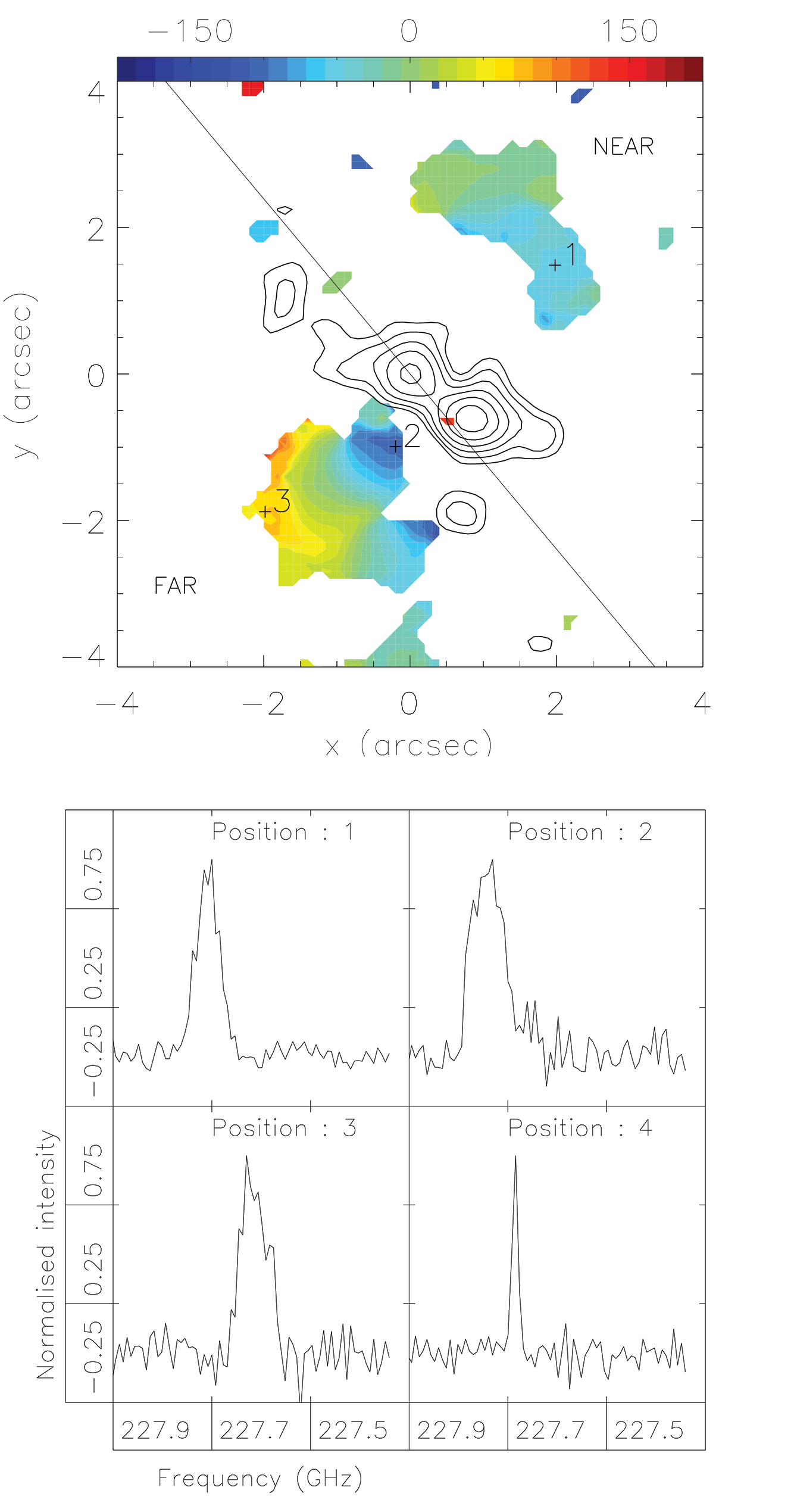}

\caption{Top: observed CO J:2-1 velocity field of the inner region is shown in colour following the colour bar (units of km/s). Black contours show the ALMA 230 GHz continuum map.
Bottom: example CO J:2-1 spectra of three distinct positions (1 to 3) in the inner SE feature as identified in the left panel, plus, for comparison, the spectrum of a fourth position (4) located outside the inner region. }
\label{secomp_spectra}
\end{figure}

The velocity dispersion (moment 2) map is shown in the rightmost panel of Fig. \ref{alma_moments}:  the highest values are observed in the inner region, with the highest dispersion found in the SE component. Example spectra of this region can be found in Fig. \ref{secomp_spectra}.

The pv-diagram taken along a slit of PA $-50 \degree$ (Fig. \ref{pvd_alma_secomp}), centred on the SE feature, shows a clear gradient in velocity along the SE feature. A second pv-diagram extracted along the minor axis, centred on the galaxy centre, using a natural weighted, 4-channel averaged image ($10$ km/s) is shown in Fig. \ref{pvd_alma_secomp}. The clear gradient of the SE component seems to follow a PA close to the minor axis, and to the PA of the nuclear bar (Fig. \ref{overlay_hst}). The unusual kinematics of this inner region is addressed in Sect. \ref{subsect:nuclear_bar}, where we argue that it is a possible result of the interaction between the large scale bar and the nuclear bar.

A map of the 230 GHz continuum (close to the CO J:2-1 emission line, Fig. \ref{secomp_spectra}) shows three components whose positions closely corresponds to the nucleus and radio lobes observed in the VLA 8.4 GHz data \citep[See Table 2 and 3 of ][]{koss2015}. The 230~GHz total fluxes are 0.2~mJy for the nuclear source, 0.31 mJy for the SW source, and 0.08 mJy for the NE source. Given the match in position of the three components observed in the milimetre continuum with the radio lobes observed in the VLA 4.9 - 8.4 GHz emission, and that the observed 230 GHz fluxes are in close agreement with the extrapolation of the 8.4 GHz fluxed and the 4.9 to 8.4 GHz spectral indices, we conclude these are the same components as those observed and analised in detail by \citet{koss2015}.

\begin{figure}
\centering
\includegraphics[width=0.5\textwidth]{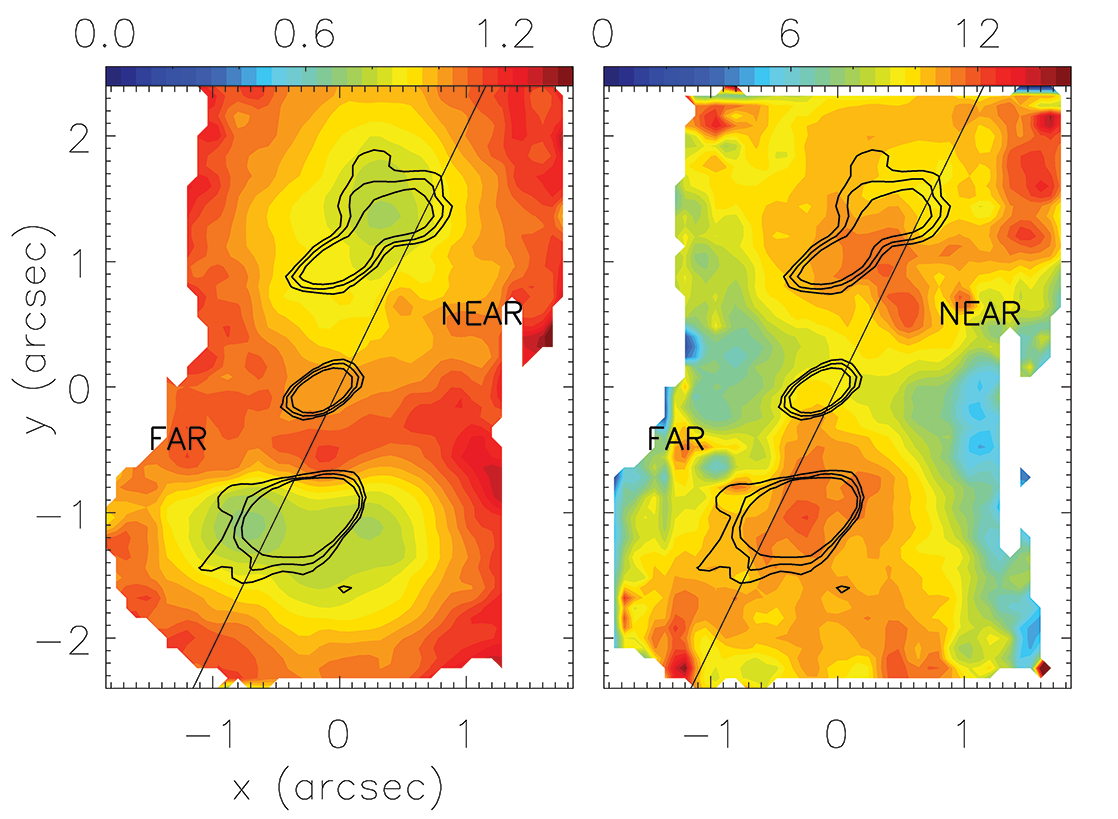}

\caption{Left: $[SII] \lambda 6716 / \lambda 6731$ line ratio map. Right: [OIII]/H$\beta$ line ratio map. Black contours correspond to the VLA 8.4 GHz map. The major axis of the galaxy (PA $= 40\degree$) is shown with the black line.}
\label{lineratios_sii_oiiihb}
\end{figure}


\section{Discussion}
\label{sect:discussion}

The main nuclear radio features of the galaxy are the nuclear component, with a flat spectrum, and two hotspots, NE ($\sim$2\arcsec) and SW ($\sim$1\arcsec) from the nucleus, with a steeper spectra, indicative of a two-sided jet. The larger flux in the SW lobe is attributed to Doppler boosting by \citet{koss2015}, assuming that the NE component is receding and the SW is approaching. There are no other previous studies that allude to the jet position in the galaxy in greater detail. \\
Based on the kinematics presented here, the edge-on maser disk with extension $1 \sim $pc in PA perpendicular to both the jet axis and the major axis of the galaxy, and the relatively low inclination of the galaxy disk (as compared to the maser disk), we interpret the jets as being launched into the disk of the galaxy. The SW  lobe may represent the point where the radio jet leaves the disk of the galaxy. This scenario implies a maximum interaction between the radio jet and the dense gas in the galaxy disk. \\

 The S-shaped morphology and the possible origin of these arms have been discussed in detail in \citet{cooke2000} and \citet{maksym2017}, where radial jet motion, entrainment of patterns in the ISM and accretion disk precession have been suggested as possibilities.
Maps of gas excitation (as traced by the [OIII]/H$\beta$ ratio) and density (high ratios of $[SII] \lambda 6716 / \lambda 6731$ denote higher densities) are shown in Fig. \ref{lineratios_sii_oiiihb}. The region of higher gas excitation appears to be biconical, centred on the nucleus, in PA $\sim$50$\degree$, and with an opening angle of 45\arcdeg. This conical morphology is most obvious to the SW. Similar morphologies can be observed in the line ratio maps of [OIII]/H$\alpha$, and [SII]/H$\alpha$, where the ratio is lower inside the bicone, and higher outside, which indicates possible shocks and lower photoionisation outside the cones. However, giving that the profiles of the H$\alpha$ line is blended with the [NII] doublet in the areas where the kinematics are more complex, the maps produced are not as reliable as that of [OIII]/H$\beta$. Despite this, the general biconical shape, with high excitation inside the cones, is maintained but the specific values for the line ratios can change depending on the Gaussian fit or spectral window used to obtain the flux of the lines. 
Similar results, with Seyfert-like emission inside the biconical shape that encloses the S-shaped arms and LINER-like emission outside the s-shaped arms, has been found by \citet{maksym2017}. 
It is thus likely that an ionisation cone is present in the galaxy along a PA $\sim 50 \degree$ and with an inclination similar to the galaxy inclination. However, the gas illuminated in this ionisation cone has kinematics which is most likely dominated by radio jet interactions with the gas in the FOV of our GMOS/IFU data. As shown in \citet{cooke2000} a high-excitation gas region extends to $\sim 20 \arcsec$ and thus a larger FOV will help to constrain the presence and characteristics of this ionisation cone.

\subsection{Outflows}

The masking and multi-component Gaussian fit to the [OIII] emission line discussed in Sect. \ref{sect:results} shows four areas of interest that we have labeled 'O1', 'O2', 'O3' and 'O4' (Fig. \ref{gmos_multicomp_mom1}).
The 'O1' region is $1\farcs8$ NE of the nucleus, near the NE radio lobe. The spectrum of this region shows a wide profile ($\sim 500$ km/s) which we have fitted with the broad redshifted component. However, both the broad redshifted component and the narrow component, which follows the expectations of rotation in the galaxy disk, have a similar velocity, and thus we do not consider a flow of gas in this region. The  $[SII] \lambda 6716 / \lambda 6731$ line ratio map shows an area of larger electron density near the 'O1' region (Fig. \ref{lineratios_sii_oiiihb}), which can indicate that shocks are being produced by the interaction between the radio lobe and the ionised gas. \\
The 'O2' region, near the NE radio lobe, shows a wide profile that seems to have multiple components present (Fig. \ref{multicomp_spectra}) and is fitted by a narrow component and broad redshifted component; this broad component shows velocities redshifted $\sim 180$ km/s from systemic velocity. We consider this to be an outflow produced by the radio jet in the plane of the disk.
The 'O3' region, near the SW radio lobe, shows a profile with a clear broad, redshifted wing. We fitted this profile with a narrow component that seems to be following the rotation of the disk, and a broad redshifted component, this broad component is redshifted $\sim 200$ km/s from systemic velocity. We consider this redshifted emission to be an outflow produced by the radio jet in the plane of the disk. \\
The 'O4' area shows the widest profiles ($\sim 700$ km/s) observed in our FOV and clearly shows three different components, which we fitted with three Gaussian profiles:  narrow, broad redshifted and  broad blueshifted components. This area extends between the nuclear and SW radio components, and along the equatorial region (that is, perpendicular to the radio jet  and  nuclear maser disk rotation axis). This region shows a lower electron density (Fig. \ref{lineratios_sii_oiiihb}) and high-velocity blueshifted emission is observed along the area, while redshifted emission is observed on a more concentrated subsection of the area, closer to the SW radio lobe. We interpret this to be an equatorial (w.r.t. the central engine) outflow.\\
In the following sections we will discuss the NE ('O2') and SW ('O3') jet-driven outflows, and the equatorial outflow ('O4') in more detail.

\subsubsection{Jet driven outflow}
\label{subsect:jetoutflow}

To estimate the ionised gas outflow rate we estimate the mass of the gas and the velocity of the outflow, following \citet{lena2015}. The gas mass is given by

\begin{equation}
 M_{g} = N_{e} m_{p} V f
\label{eq1_mass}
\end{equation}

where $N_{e}$ is the electron density, $m_{p}$ is the proton mass, $V$ is the volume considered and $f$ is the filling factor, which can be estimated by 

\begin{equation}
 L_{H_{\alpha}} \sim f N_{e}^{2} j_{H_{\alpha}}(T) V 
 \label{eq2_fillingfactor}
\end{equation}

where $L_{H_{\alpha}}$ is the $H_{\alpha}$ luminosity emitted by the volume $V$, and    $j_{H_{\alpha}} = 3.534 \times 10^{-25}$ erg cm$^{-3}$ s$^{-1}$ \citep{Osterbrock1989}. Combining these two equations we obtain:

\begin{equation}
M_{g} = \frac{m_{p}L_{H_{\alpha}}}{n_{e}j_{H_{\alpha}}(T)}
\label{eq3_finalmass}
\end{equation}

To estimate the outflow rate we use an aperture of $0\farcs6$ for each component. For the 'O2' outflow, the mean $[SII] \lambda 6716 / \lambda 6731$ line ratio of the broad redshifted Gaussian component is $0.92$, which corresponds to an electron density of $740$ cm$^{-3}$, the H$\alpha$ luminosity of the same Gaussian component in this aperture is $3.4 \times 10^{40}$ erg s$^{-1}$, and the gas mass that is outflowing, assuming a luminosity distance to the galaxy of $52 $ Mpc, is $11 \times 10^{4} \; M_{\odot}$; The mean deprojected velocity in the aperture is $180$ km/s. This gives an outflow rate for 'O2' of $\dot M = 0.07$ M$_{\odot}/$yr.
For 'O3', the electron density is $980$ cm$^{-3}$, the H$\alpha$ luminosity is $3.6 \times 10^{40}$, the mass is $9 \times 10^{4}$ M$_{\odot}$. The mean deprojected velocity is $210 $ km/s. Thus, the outflow rate for  'O3' is $\dot M = 0.06$ M$_{\odot}/$yr.

\subsubsection{Equatorial outflow}
\label{subsect:eqoutflow}

The [OIII] multi-component Gaussian fits shows the presence of a strong broad blueshifted component along the equatorial region, on a strip $\sim 1\arcsec$ wide along PA $147 \degree$. The velocity map of this component is shown in Fig. \ref{gmos_multicomp_mom1} (marked as 'O4') where high blueshifted velocities can be observed in the equatorial region, perpendicular to both the galaxy disk PA and the radio jet axis. The presence of a weak redshifted component can be inferred from a redshifted wing, visible on a compact fraction of the equatorial region (Fig. \ref{gmos_multicomp_mom1}), and is not present on the entire equatorial region as the blueshifted component is. 
This distribution and the presence of a blue component in the region perpendicular to the radio jet axis indicates the presence of an equatorial outflow along PA $147 \degree$, which is in good agreement with the water maser disk PA, which is nearly perpendicular to the radio jet axis and extends for a $\sim 1$pc \citet{kondratko2008}. In this scenario the redshifted component will be behind the galaxy disk and thus will appear weaker due to obscuration in the disk, as indicated in \citet{cooke2000} the presence of a dust lane is observed in the central $0\farcs5$ of the continuum peak (HST f547m filter) along PA $\sim 115\degree$, this dust lane is also observed by \citet{pogge1997} in the HST F606W. The interaction of a strong equatorial outflow from the accretion disk with the surrounding gas can push the ionised gas outwards. 

Although the presence of outflows along the radio jet is more common \citep[e.g.][]{das2007,barbosa2009,storchi-bergmann2010}, equatorial outflows have been included in theoretical models of accretion disk winds \citep{li+ostriker2013}, and outflowing torus models \citep{honig2013,elitzur2012}, and observationally found in NGC 5929 \citep{riffel2014,riffel2015} and NGC 1386 \citep{lena2015c} where a distinct component that involves rotation and/or outflow extends to 2-3\arcsec ($\sim$ 200 pc) at either side of the nucleus, an extension similar to that found in NGC 3393.

In this scenario, and given that the galaxy is almost face-on, we assume that the blueshifted gas is in front and possibly leaving the disk and the redshifted gas is behind. 
The mean $[SII] \lambda 6716 / \lambda 6731$ line ratio of the blue component in the equatorial region is  $0.93$ which corresponds to an electron density of $720$ cm$^{-3}$.
The $H_{\alpha}$ luminosity of the region is $ 5 \times 10^{40}$ erg s$^{-1}$. From Eqn. \ref{eq3_finalmass} we obtain a mass of $M_{g} = 2 \times 10^{5} \; M_{\odot}$, the mean observed velocity  is $-420$ km/s and we consider this to be the true velocity of the gas, i.e. this outflowing gas is not in the plane of the disk, but leaving it, and approaching in the line of sight. \\
The estimated equatorial outflow rate, under these assumptions, is $\dot M = 0.24$ M$_{\odot}/$yr.
An alternative method to derive the equatorial outflow rate is to assume a hollow cylinder geometry that is expanding from the centre, with a height of $0\farcs5$. In this case the estimated outflow rate is $7$ M$_{\odot}$/yr, assuming a filling factor $f = 0.01$ \citep[following][]{riffel2015}. The differences in these two outflow rate estimates suggests the need of a filling factor closer to $\sim$0.001 or a significantly smaller height. \\

\subsection{Bar perturbations}
\label{subsect:bar_descp}

To understand the role of the bar-induced perturbations to the molecular gas kinematic we have applied the harmonic decomposition formalism described in \citet{sfdz1997} and \citet{wong2004}. It is important to remark that this formalism is based on linear epicycle theory, and thus it is only valid for weak bars, as the departure from circular orbits must be small compared to the circular velocity. The line of sight velocity towards a given point in a galaxy velocity field can be decomposed in a Fourier series as: 
$$ V_{LOS} (R) = c_{0} + \sum _{j=1}^{n} [ c_{j} \cos(j\psi) + s_{j} \sin(j\psi) ]\sin i \; , $$

\noindent where $(R,\psi)$ are polar coordinates, $i$ is the inclination of the disk, $c_{0}$ corresponds to the systemic velocity ($V_{sys}$), and $j$ is  the harmonic number. The coefficients $c_{j}$ and $s_{j}$ are a function of the circular velocity ($V_{c}$), the bar pattern speed ($\Omega_{p}$), ellipticity of the potential ($\epsilon$) and the bar viewing angle ($\theta_{obs}$), which corresponds to the bar PA from the minor axis of the galaxy disk \citep[see e.g. Fig. 2 of][for a definition of this angle]{wong2004}. A bar creates a bisymmetric gravitational potential which has a predominant $m=2$ Fourier component, and thus we only consider the harmonics $j=m-1$ and $j=m+1$ \citep{schoenmakers1997}.
For the circular velocity ($V_{c}$) we used the value obtained from the best fit exponential disk potential (Fig. \ref{alma_model_expdisk} and Sect. \ref{subsect:molecular_gas}).

\begin{figure*}
  \centering
  \includegraphics[width=0.99\textwidth]{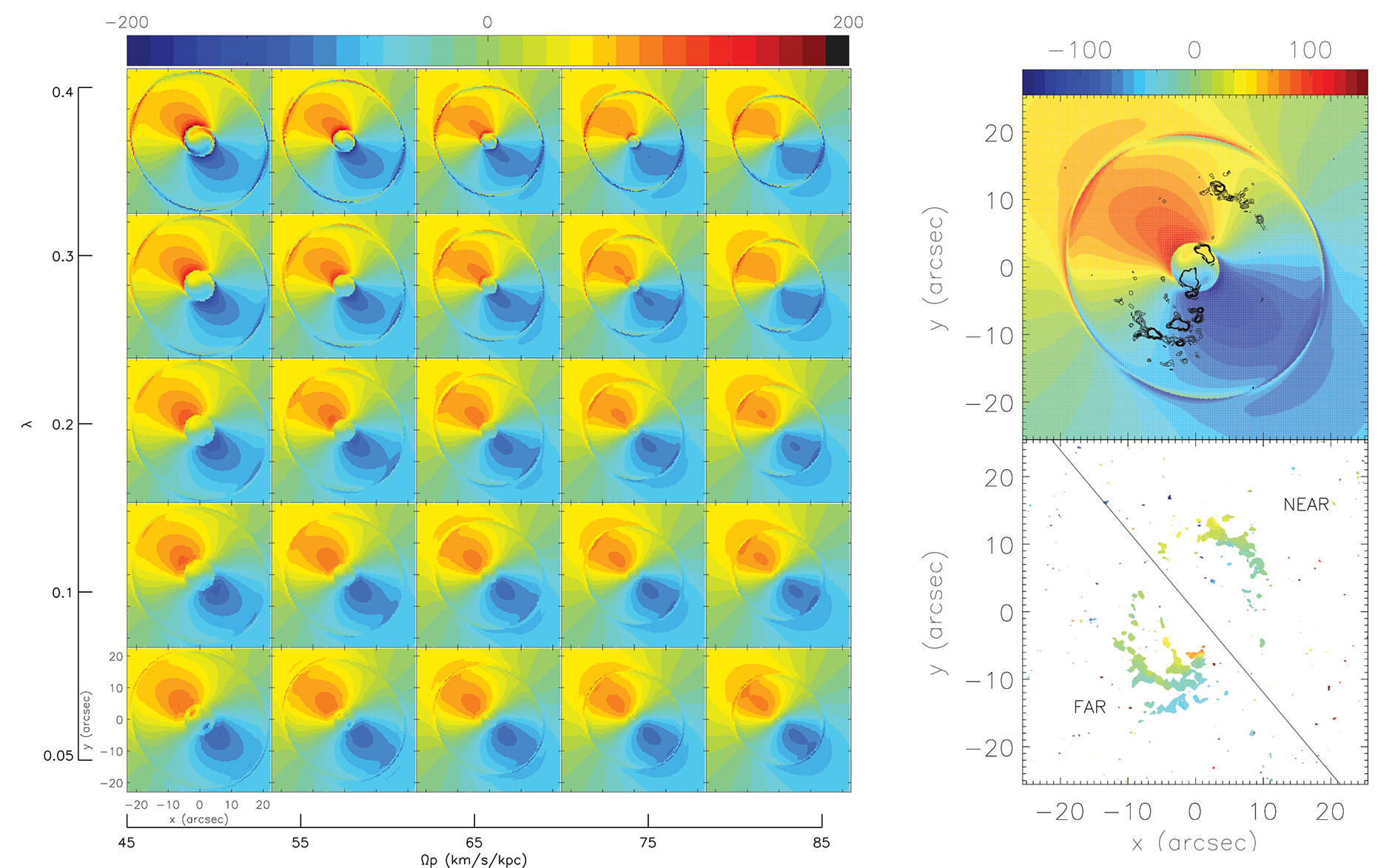}
  \caption{Left: Velocity fields resulting from the bar perturbation model described in Sect. \ref{subsect:bar_descp}, when varying $\Omega_{p}$ (x-axis, values from 45 to 85 km/s/kpc, with a 10 km/s/kpc step), and $\lambda$ (y-axis, values of 0.05, 0.1, 0.2, 0.3, and 0.4). All models use the intrinsic rotation curve derived from the best fit exponential disk model with parameters as explained in Sect. \ref{subsect:molecular_gas}, and the bar parameters used were those of the large-scale bar: $PA_{bar} = 160\degree$ and $\epsilon=0.15$. All panels follow the colour bar shown at the top (units of km/s). Top Right: velocity field of the best-fit bar perturbation model (see text) to the large scale (outside the grey circle in Fig. \ref{alma_moments}) CO velocity field following the color bar above the panel (units of km/s). The values of the exponential disk parameters, disk PA and inclination, were fixed to the values outlined in Fig. \ref{alma_model_expdisk}, and the bar PA ($PA_{bar} = 160\degree$) and ellipticity ($\epsilon = 0.15$) were set to the values of the large-scale bar. The best fit values for $\lambda$ and $\Omega_{p}$ are 0.1 and 54 km/s/kpc, respectively. Overlaid black contours show the integrated intensity (moment 0) of  CO J:2-1. Bottom Right: residual (observed - model) velocity map for the bar perturbation model following the color bar above the panel (units of km/s): only the large scale velocity residuals are shown.
 }
\label{bar_models}
\end{figure*}

A constant damping term ($\lambda$), simulating a radial frictional force, is introduced \citep[e.g.][]{lindblad+lindblad1994,wada1994} to avoid sudden changes in orbit axes and thus multiple orbits at a given point. Usual values for this parameter are in the range $0.05 < \lambda < 0.4$.

For the case of NGC 3393, $\Omega_{p}$ and $\lambda$ are the parameters with the largest uncertainties. We thus build a library of models with different parameter combinations, a section of which is shown in Fig. \ref{bar_models}, where the two parameters are varied over $ 0.01 < \lambda < 0.2$ and $ 25 < \Omega_{p} < 85$ km/s/kpc. The primary effect of changing $\Omega_{p}$ is the variation in the radii of the resonances, but it also effects the magnitude of twists in the isophotes. The effect of increasing $\lambda$ is to smoothen the sudden twists seen near the resonances. \\

We attempted to fit the full CO velocity field to this perturbation theory model, with $\Omega_{p}$ and $\lambda$ as free parameters, and the perturbations terms set by the large-scale bar PA and ellipticity. However, we could not find any suitable set of parameters (even if the bar PA was varied) which could match both the outer CO velocities and the highly perturbed velocities in the SE inner CO component. We are thus forced to conclude that a single $m=2$ (i.e. bar) mode is unable to explain the complex molecular kinematics seen in NGC 3393. The remaining option is thus to attempt to separately fit perturbations in the large-scale molecular kinematics (driven by the large-scale bar) and in the inner molecular kinematics (driven by the nuclear bar), which we do in the following subsections.

\subsubsection{Large-scale bar}
\label{subsect:large_scale_bar}

To obtain the (large-scale) bar perturbation model which best fits the outer CO kinematics we fit the observed CO velocity field (outside the grey circle in Fig. \ref{alma_moments}) to the predictions of our bar-perturbation models by using the same least-square minimization routine as explained above, in order to obtain the best-fitted parameters. We fix the exponential disk model parameters to those obtained above. We also fix the ellipticity of the potential ($\epsilon = 0.15$) and the bar PA ($PA_{bar} = 160 \degree$) to the values obtained for the large-scale bar by \citet{jungwiert1997}. The free parameters are thus the damping term ($\lambda$) and the bar pattern speed ($\Omega_{p}$).

The resulting best-fit model obtained and its velocity residuals are shown in Fig. \ref{bar_models}. The r.m.s. of this residual map is lower than that obtained when only the pure rotational model of the exponential disk is used, though the difference is not large. To further test the suitability of the best fit model we compare the model with data extracted along the minor axis in the pv-diagram shown in Fig. \ref{pvd_alma_secomp}. In the outer region, both model and data are close to zero as expected along the minor axis, however the model does not fit the data in the inner region, where the large velocity gradient of the SE feature is observed. This exercise shows that a bar perturbation model could maintain velocities similar to the exponential disk model in the outer region while having a different PA and velocity distribution inside the resonance radius.\\

The resonances observed in the best fitted model correspond to the ILR (at r = 3.7\arcsec), OLR (r = 20\arcsec), an CR (r = 13.5\arcsec) which is in good agreement with the length of the large scale bar (SMA $\sim$ 13\arcsec according to \citet{alonso-herrero1998,lasker2016}). The ILR encloses the nuclear region, including both SE and SW features, and the OLR encloses the outer distribution of molecular gas. These two resonances could help explain the difference in PA of the ALMA and GMOS data compared to the large-scale kinematics position angle \citep[PA = 68$\degree$, according to][]{cooke2000}. It is also interesting to note that the kinematics enclosed in the ILR resemble the observed stellar velocity map, specially the S-shaped zero velocity line (Fig. \ref{stellar_kinematics}). \\

The primary limitation in the analysis above is the sparse filling factor of CO velocities over the FOV. Deeper integrations with ALMA are thus crucially required. Alternatively, deep observations of the ionised gas kinematics over the full galaxy (using e.g., VLT/MUSE) are required. Indeed, the latter have been recently obtained by another team.

\subsubsection{Nuclear bar}
\label{subsect:nuclear_bar}

\begin{figure}
\centering
\includegraphics[width=0.5\textwidth]{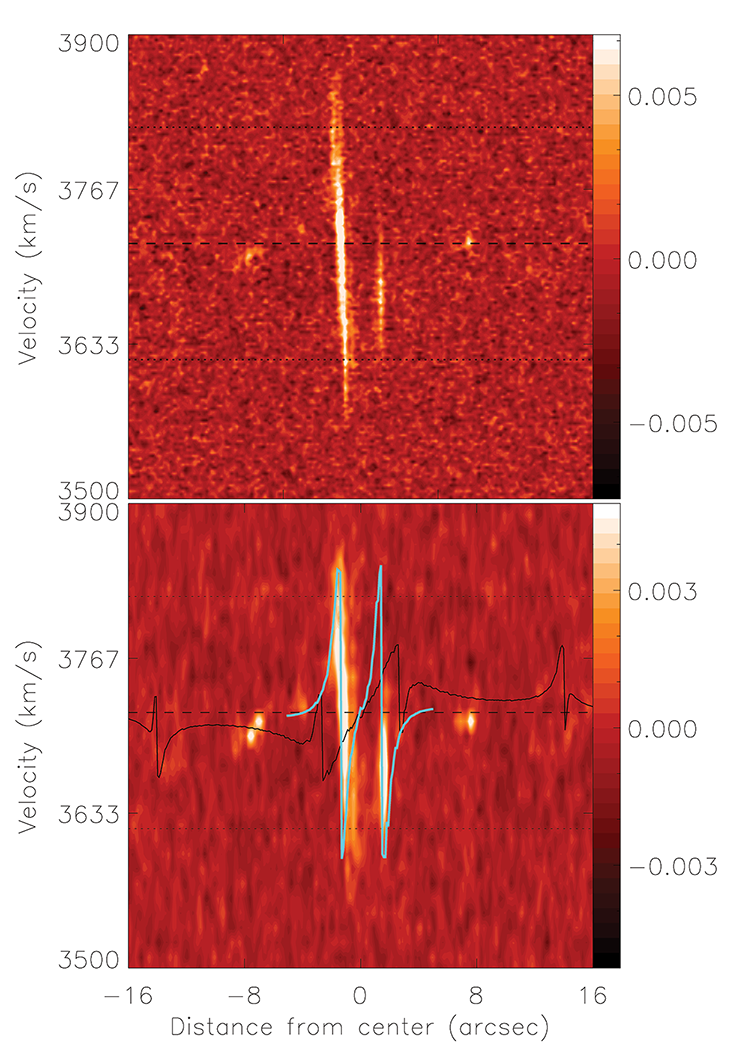}

  \caption{Top: pv-diagram centred on the SE component in the CO maps, extracted along a PA of $-50 \degree$. Bottom: pv-diagram extracted from a  natural weighted, 4 channel averaged cube with 10 km/s spectral resolution, with a slit centred on the galaxy nucleus and extracted along the minor axis. The black line shows the prediction of the large-scale bar perturbation model described in Sect. \ref{subsect:large_scale_bar} (see Fig. \ref{bar_models}) and the blue line shows the prediction of the nuclear bar model. Color bar units are Jy/beam.
}
\label{pvd_alma_secomp}
\end{figure}

\begin{figure}
 \centering
  \includegraphics[width=0.5\textwidth]{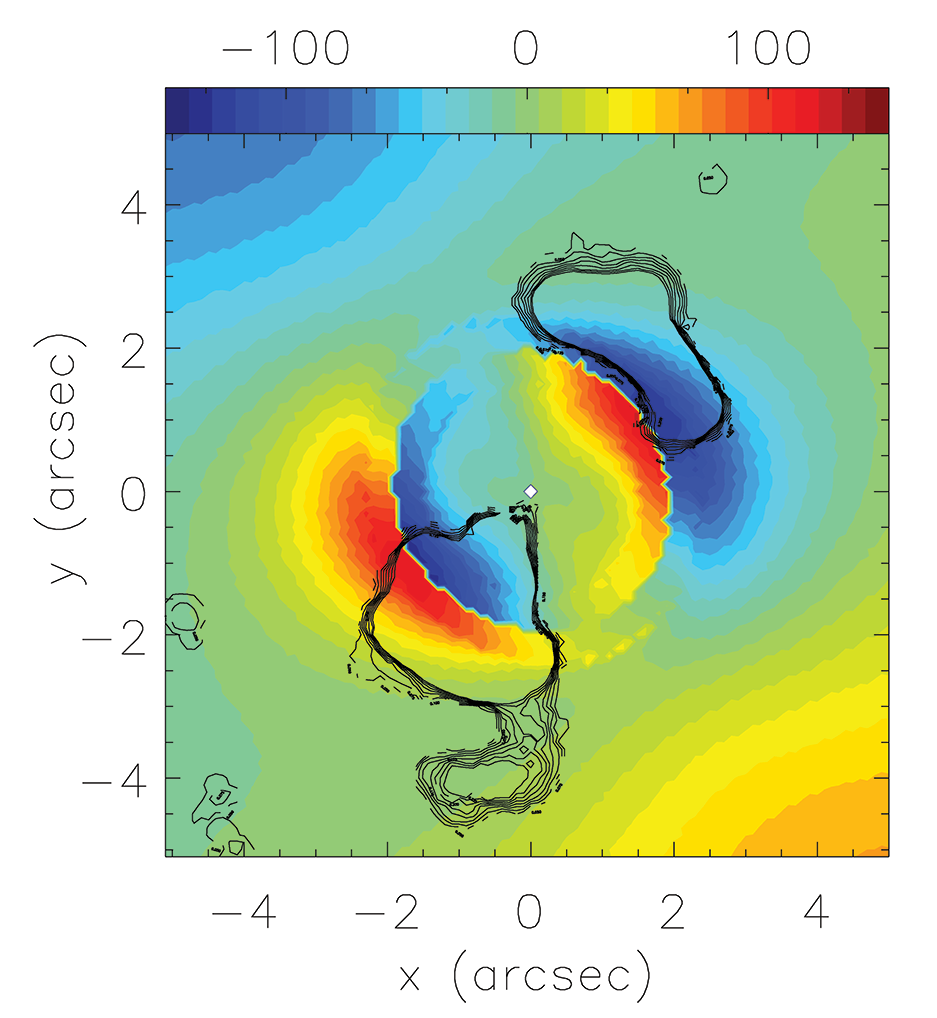}
  \caption{Velocity field of the best fit bar perturbation model for the nuclear bar shown in colour following the colour bar above the panel (units of km/s). Here the fit was made only to the inner region (inside the grey circle in Fig. \ref{alma_moments}) CO velocity field. The values of the exponential disk, disk PA and inclination, were fixed to the values outlined in Fig. \ref{alma_model_expdisk}. The best fit values obtained were $\theta_{obs} = -85\degree$, $\epsilon = 0.35$, $\lambda = 0.1$ and $\Omega_{p} = 73$ km/s/kpc.}
\label{chosen_innerbar_model}
\end{figure}

The presence of an additional nuclear bar has been suggested by NIR imaging \citep{alonso-herrero1998} and by light distribution modelling from HST imaging \citep{lasker2016}. The extension of (SMA) this nuclear bar is $\sim 2\arcsec$, and it is offset from the large scale bar by $10\degree - 20 \degree$. \\
Considering that both the PA and extension of the inner features of our CO J:2-1 data agree with those of the nuclear bar, we build bar perturbation models for the nuclear bar (Fig. \ref{chosen_innerbar_model}), assuming that the nuclear bar is decoupled from the large-scale bar and thus has a larger pattern speed. The disk parameters (exponential disk parameters, disk PA and inclination) were kept fixed to the values used above. Typical ranges were assumed for the free parameters: $0.15 < \lambda < 0.4$, $ 0.1 < \epsilon < 0.8$, $ 10 < \Omega_{p} < 200 $ km/s/kpc. The best fitted values obtained were: $\lambda = 0.1 $, $\Omega_{p} = 73$ km/s/kpc and $\epsilon = 0.35$, and $\theta_{obs} = -85\degree$. The latter value implies that the large-scale and nuclear bars are almost aligned, and that the nuclear bar is completely embedded in the large-scale bar; which is consistent with the results of \citet{alonso-herrero1998}, where the PA difference between both bars is in the range $10\degree-20 \degree$.
The resulting model is overlaid on the PV-diagram along the minor axis (Fig. \ref{pvd_alma_secomp}), where it can be observed that it follows the same gradient of the inner SE feature and the Keplerian-like fall-off, while it is also close to the gradient of the inner NW feature. \\

Considering that we can reproduce a similar gradient as that observed in the inner SE feature, it is possible that an interaction of the large-scale bar and the nuclear bar exists. If these components present kinematical coupling, they can share a resonance, usually the ILR for the large-scale bar coincides with the CR of the nuclear bar. If this is the case for NGC 3393 it is possible that the molecular gas observed is near the ILR of the large-scale bar where it can accumulate into rings. However, the presence of an inner bar could, in principle, allow the gas to overcome this limit and continue to flow to the central regions \citep{shlosman1989}. 

This simple toy model indicates that it is feasible that a large-scale and nuclear bar interaction can produce a feature similar to the one observed in the inner region of our ALMA data. However, the kinematical coupling between both bars and the consequent complex modelling of its effect is beyond the scope of this paper. \\

\subsection{SMBH}

Evidence for a secondary SMBH has been presented by \citet{fabbiano2011}. This BH is located $0\farcs 6$ SW from the nucleus (Fig. \ref{onecomp_gh_gmos}). As it can be seen in the moment 1 maps of the ionised emission lines, there is no clear kinematical component connected to the position of this posited secondary BH. However, the kinematics of the ionised gas are deeply perturbed by the radio jet and thus any kinematical signature of a secondary SMBH can be easily lost. \\
An alternative explanation to the unusual gradient observed in the SE component of the molecular gas might be linked to this posited secondary SMBH. A pv-diagram along a PA of $50 \degree$ centred on the feature (Fig. \ref{pvd_alma_secomp}) shows a mostly smooth gradient that goes from $\sim - 100$ km/s to $\sim 100$ km/s, which could indicate a nuclear disk related to the secondary SMBH. However, the kinematic and morphological centre of the SE feature does not corresponds to the posited position of the secondary BH. Alternatively, the inner CO emission (i.e. both the SE and NW inner components) could be centred around the secondary SMBH; although we cannot rule out this case, it would require a very special geometry. Simpler explanations, such as the nuclear bar perturbations are thus favoured by us.  
The posited secondary SMBH is located in between the nuclear and the SW source, and while there is no clear detection of another source in that position, the presence of another continuum source here cannot be discarded unequivocally. Thus, while we find no evidence supporting the existence of this secondary SMBH, it is important to note that both the observational and interpretive constraints are not strong enough to disprove the presence of the secondary SMBH.

If we assume that the equatorial outflow is axisymmetric (though we have only detected the blueshifted gas in this outflow) then the mass outflow rate for the equatorial outflow can reach $\dot M \sim 0.6$ M$_{\odot}/$yr. The outflow rates presented here are a lower limit, as we only consider the ionised gas. \\ 
To contrast the outflow rates estimated here, we now estimate the inflow rate necessary to feed the SMBH. The bolometric luminosity estimated using the 2-10 keV luminosity is $2.4 \times 10^{44}-7.6 \times 10^{44}$ erg/s \citep{kondratko2008}. Assuming a standard accretion efficiency of $10\%$ the accretion mass rate required is $\dot M = 0.04-0.1$ M$_{\odot}/$yr, a factor $\sim 8$ lower than our estimated outflow rate, which is not unusual in nearby galaxies \citep{barbosa2009,muller-sanchez2011}. \\
The difference between inflow and outflow rates can indicate that the outflowing gas does not originate from close to the central engine, but is circumnuclear ISM gas that is being pushed away by the radio jet. While we do not find direct evidence of inflows in the ionised gas, if the SE component of the observed CO J:2-1 molecular gas was inflowing with a velocity of 10 km/s the necessary inflow rate could be achieved.
The total CO mass of the SE feature is M(H$_{2}$) = 5.4 $\times 10^{7}$ M$_{\odot}$, assuming $\alpha_{CO} = 4.6 $ M$_{\odot}$ (K km s$^{-1}$ pc$^{2}$)$^{-1}$. If the all the molecular mass enclosed in the SE feature was inflowing with a velocity of 10 km/s, the potential accretion rate produced would be 0.32 M$_{\odot}$/yr. While we do not find direct evidence that the molecular gas is inflowing, a velocity of 10 km/s would fall under the detection limit of our analysis, and the interaction between the large scale and nuclear bar can make this inflow possible.


\section{Conclusions}
\label{sect:conclusions}

We have analysed the kinematics of the stars, ionised gas, and molecular gas in the central kpcs of the Seyfert 2 galaxy NGC 3393 using optical integral field observations from GEMINI-GMOS/IFU and ALMA. From our detailed analysis of these data we conclude that:

- NGC 3393 presents a complex multi-phase ISM, with strong interactions between the nuclear radio jet and the ionised gas produces complex kinematics. We have found that it is necessary to fit the emission line profiles observed with multiple Gaussian components. We identify three components in the ionised gas, which we refer to as the narrow, broad redshifted, and broad blueshifted components.

- The narrow ionised gas component has a low velocity dispersion ($\sigma < 115$ km/s) and, more or less, follows pure rotation in the galaxy disk; nevertheless this component is perturbed in the regions near the radio lobes.

- The broad redshifted component ($ 115 < \sigma < 230 $ km/s) can be observed in regions near the radio lobes. We identify two outflows in this component named as 'O2' and 'O3'.  'O2' seems associated to the NE radio lobe, while 'O3' is near the SW radio lobe. The 'O2' outflow is redshifted on the far side of the galaxy which can indicate gas being pushed away by the radio jet.  As the SW radio lobe appears to be the approaching component of the radio jet, it is possible that this jet is starting to leave the galaxy in the region of the 'O3' outflow, and thus the redshifted gas observed can be gas in the disk, being pushed away from the line of sight by the radio jet, or if the jet remains in the disk the outflow is produced by the gas being pushed by the jet inside the disk plane.

- The broad blueshifted component ($ 115 < \sigma < 230 $ km/s) presents large blueshifted velocities distributed along the equatorial region, perpendicular to the radio jet axis. A weaker redshifted wing it is also visible in the same region. We interpret this component as being part of an equatorial outflow, expanding perpendicular to the radio jet axis, and  whose emission is attenuated by dust in the plane of the  galaxy.

- From the measured velocities, $H_{\alpha}$ fluxes, and electron densities, of the outflowing components, we estimate a total outflow rate of $\dot M \sim 0.13$ M$_{\odot}$/yr for the jet driven outflows, and $\dot M \sim 0.24$ M$_{\odot}$/yr for the equatorial outflow.
If we consider a symmetric component for the equatorial outflow, the total outflow rate can reach $\dot M \sim 0.6$ M$_{\odot}$/yr for the ionised gas. This outflow rate is $\sim 8$ times larger than the accretion rate necessary to fuel the AGN. While we found no direct evidence for gas inflows, we note that the necessary inflow rate can be provided if the SE component of the CO J:2-1 emission is inflowing at a velocity of $\sim 10$ km/s, a velocity which would be close to the detection limit of our observations and analysis.

- We were forced to analyse the kinematics for the CO J:2-1 emission separately for two regions, an inner region within $5''$ of the nucleus, and the region outside said circle, since we could not find a global model which could fit both regions together. We do not detect CO J:2-1 emission at either position of the SMBH or at the position of the posited secondary SMBH. 

- The outer region of the CO J:2-1 emission seems to trace the rotation in the outer disk, and can be fitted with an exponential disk rotation model, though obvious residuals remain. To understand the role of the large-scale bar in the kinematics observed on the CO J:2-1 emission we applied the harmonic decomposition formalism to the CO velocity field. Specifically, we fitted a bar-perturbation model to the outer region of our CO J:2-1 velocity field. We found, over a range of different $\Omega_{p}$ and $\lambda$, the presence of a resonance between the inner and outer region, and a resonance that encloses the outer region of the CO emission. These resonances could explain the difference in PA found in the ALMA and GEMINI-GMOS/IFU data compared to the large-scale kinematics observed by \citet{cooke2000}, and the observed distribution of CO J:2-1 emission.
This model, however, does not fit the observed CO kinematics of the inner region.

- The inner region of the CO J:2-1 emission presents highly disturbed kinematics, with the presence of an off-nuclear velocity gradient centred in the SE component. We found this gradient can not be explained by the large-scale bar model, nor by the presence of the posited secondary SMBH or any disk related to it. We fitted a second bar perturbation model based on the parameters of the nuclear bar and found a good fit to the inner region kinematics. This toy model indicates
that the kinematics observed in the inner region of the CO J:2-1 emission can be attributed solely, or at least dominantly, to perturbation by the nuclear bar, together with interactions between the  large-scale and nuclear bars.

\section*{Acknowledgements}
This work was supported by CONICYT PhD fellowship 2015-21151141.
NN acknowledges Fondecyt 1171506, Conicyt ALMA 3114007, and BASAL PFB-06/2006.
This paper makes use of the following ALMA data: ADS/JAO.ALMA\# 2015.1.00086.S. ALMA is a partnership of ESO (representing its member states), NSF (USA) and NINS (Japan), together with NRC (Canada), NSC and ASIAA (Taiwan), and KASI (Republic of Korea), in cooperation with the Republic of Chile. The Joint ALMA Observatory is operated by ESO, AUI/NRAO and NAOJ.
Based on observations obtained at the Gemini Observatory, which is operated by the Association of Universities for Research in Astronomy, Inc., under a cooperative agreement with the NSF on behalf of the Gemini partnership: the National Science Foundation (United States), the National Research Council (Canada), CONICYT (Chile), Ministerio de Ciencia, Tecnolog\'ia e Innovaci\'on Productiva (Argentina), and Minist\'erio da Ci\^{e}ncia, Tecnologia e Inova\c{c}\~{a}o (Brazil).


\bibliographystyle{mnras}
\bibliography{refs}





\bsp	
\label{lastpage}
\end{document}